\documentclass[twocolumn,aps,pre,amsmath,amssymb,groupedaddress]{revtex4-2}
\usepackage{amsmath}
\usepackage{amssymb}
\usepackage{graphicx}

\makeatletter
\usepackage{units}

\usepackage{verbatim}
\usepackage{color}

\usepackage{dcolumn}
\usepackage{bm}
\usepackage{amsfonts}
\usepackage{pifont}
\usepackage{amsfonts}

\newcommand{\beq}{\begin{eqnarray}}
\newcommand{\eeq}{\end{eqnarray}}

\makeatother

\begin{document}
\title{The Physics of Self-Rolling Viruses}
\author{Pedro A. Soria Ruiz$^{1}$, Falko Ziebert$^{1,2}$, and Igor M. Kuli\'{c}$^{3,4}$}
\affiliation{$^{1}$Institute for Theoretical Physics, Heidelberg University, D-69120 Heidelberg, Germany\\
$^{2}$BioQuant, Heidelberg University, D-69120 Heidelberg, Germany\\
$^{3}$Institut Charles Sadron UPR22-CNRS, F-67034 Strasbourg, France\\
$^{4}$Institute Theory of Polymers, Leibniz-Institute of Polymer Research,
D-01069 Dresden, Germany
}

\date{\today}
\begin{abstract}
Viruses are right at the interface of inanimate matter and life.
However, recent experiments [T. Sakai, et al., J.~Virol.~{\bf 92}, e01522-17 (2018)]
have shown that some influenza strains can actively roll on glycan-covered surfaces.
In a previous  letter [F. Ziebert and I. M. Kuli\'{c}, Phys. Rev. Lett. {\bf 126}, 218101 (2021)]
we suggested this to be a form of viral surface metabolism: a collection of spike proteins that attach to and cut the glycans
act as a self-organized mechano-chemical motor. Here we study in more depth the physics of the emergent self-rolling states.  
We give scaling arguments how the motion arises, substantiated by a detailed analytical theory
that yields the full 
torque-angular velocity relation of the self-organized motor.
Stochastic Gillespie simulations are used to validate the theory and to quantify 
stochastic effects like virus detachment and reversals of its direction.
Finally, we also cross-check several approximations made previously and show that the proposed mechanism
is very robust. All these results point together to the statistical inevitability of viral rolling
in presence of enzymatic activity. 

\end{abstract}
\maketitle

\section{Introduction}

One of humanity's greatest inventions is the wheel and 
technological revolutions were carried by its ''motorization''.
Being technologically so indispensable, we might ask for the wheel's utility
in biology \cite{RichardDawkins}. On the macroscale examples are scarce, 
yet in the micro realm passive rolling, e.g.~of white blood cells \cite{Hammer_WBC}
or malaria-infected red blood cells \cite{Cooke}, occur in shear flow and are important for the 
functioning of the immune response and the traveling of the parasite through our body, respectively.
Active self-propelled rolling was unknown for a long time.
But surprisingly, the motorized wheel was rolling also in nature for ages:
our old molecular adversary -- the influenza virus -- apparently is able to use its whole body 
as a chemically driven monowheel that actively rolls on our lung cells' surfaces by catalytically
hydrolyzing sugars sticking out from the cell membranes. 
This surprising (and maybe even alarming) phenomenon of active virus surface-rolling
has been  demonstrated first by Sakai et al.~\cite{Sakai_Saito_IVA,Sakai_Saito_IVC}
and interpreted as a Brownian-ratchet-like effect. It has been also observed indirectly \cite{Guo_deHaan}
and is discussed now as an important pathway helping the virus to cross and navigate the mucus \cite{deVries,deVries2}. 

The underlying physical mechanism -- different from the classical Brownian burnt bridge model
\cite{Blumen,Krapivsky} --  has been proposed recently in \cite{virusPRL}, 
 where we outlined elements of a model which we elaborate deeper in the present work. 
The initial model appears to have left parts of the molecular motor community in slight
disbelief \cite{PhysicsFOCUS} whether the mechanism could actually work as described. 
Here we explain the robustness and inevitability of the rolling state as proposed earlier 
and verify approximations made in \cite{virusPRL}
against more detailed analytical calculations and stochastic simulations. 
In addition we develop a scaling view on the mechanism. 
We study stochastic effects such as reversals of direction and virus detachment.
And finally, we clarify how the directional stability and processivity physically emerges by pinpointing an internal ``mechano-chemical flywheel'' -- a 
long-living internal polarization mode with directional memory -- 
that allows for highly persistent rolling to occur in spite of large external noise 
and at zero Reynolds number.

The paper is organized as follows: in section \ref{basic} we describe
the basic ingredients, from reaction kinetics to force balance. Section
\ref{steady} then goes on to explain on an intuitive level why the virus actually 
rolls and how its rolling steady state arises.
In section \ref{steadyanalyt} the steady rolling is studied 
via simple approximations allowing
the force-velocity relation to be analytically investigated.
Section \ref{stochastic} then describes a stochastic version of the
model which we show to be consistent with the continuum version in
the steady state. Beyond this limit, 
we also 
explore stochastic phenomena, like 
the rates of virus detachment, reversals, run lengths etc. 
Section \ref{further topics} critically scrutinizes approximations made so far 
and quantifies the dynamical persistence of the mechanism 
by studying a virus being stopped instantaneously, leading to a build-up of torque 
via the ``flywheel'' effect.
Finally, in section \ref{sec_discussion} we discuss implications of the model 
for biology/virology and we conclude with some open questions and experimental tests.

\section{Basic Model}
\label{basic}

\begin{figure}
\includegraphics[width=0.49\textwidth]{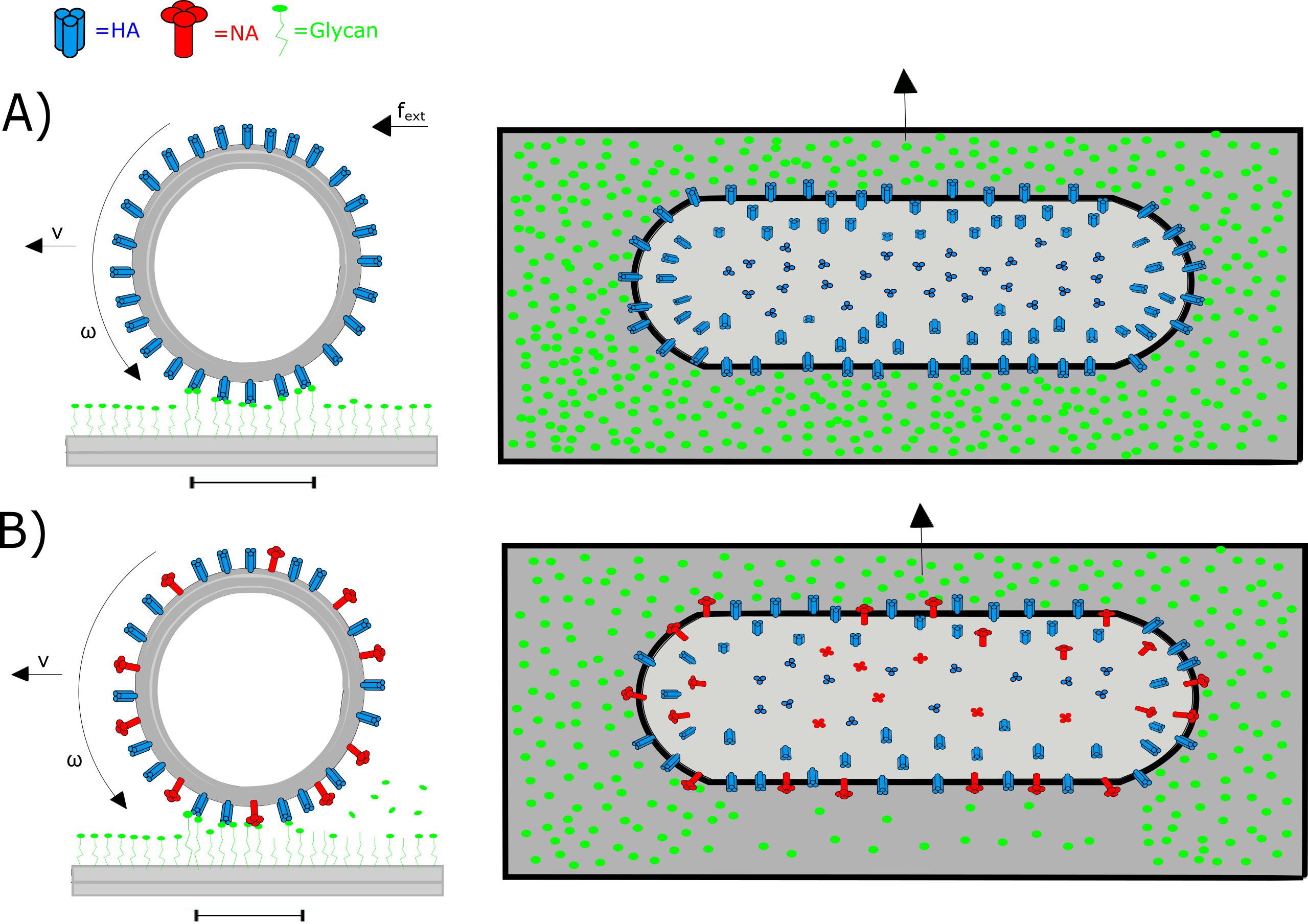}
\caption{\label{figsketch}
Influenza A has two spike proteins on its surface: hemaglutinin (HA, blue) which attaches 
to sugar residues from short surface-bound glycan chains (green) 
and neuraminidase (NA, red) which cuts the 
sugar residues such that they are no longer available for HA binding.
In a first step, we consider a virus rolling under an externally applied force
over a glycan-covered (cf.~green dots) surface,
as sketched in A) as both  a cross-section view and a top view.
The HA-glycan binding kinetics (in the contact zone as sketched in the left panel by the bar) 
will result in a friction force opposing the rolling motion.
B) In a second step, we consider the effect of NA cutting. 
The dynamic self-organization (self-polarization)
of the bound linker profile in the contact zone now allows for steady self-rolling.
The NA ``consumes'' glycans which leads to a depleted trail on the back of the virus
(see right panel). 
}
\end{figure}

We first present a mean-field description that will give much
general insight into the problem and, using several approximations,
allows for a detailed analytical treatment of the steady rolling.

The results derived here are formulated for any viral capsid cross-section 
orthogonal to the rolling direction and are equally valid for both influenza 
isoforms \cite{Badham_Rossman,Dadonaite}: 
the cylindrical (filamentous) and the spherical  virus geometry. 
The main difference between the two cases is the system size 
(total number of interacting spike proteins) and the enhanced role of fluctuations 
for spherical or smaller aspect ratio viruses. 
Stochastic effects are postponed to section \ref{stochastic},
where we also critically scrutinize the validity of the assumptions made here.

\subsection{Influenza spike proteins and their kinetics}
\label{spike_kinetics}

As one of the omnipresent molecular adversaries of mankind, 
the influenza virus (IV) and its proteins have
been extensively characterized \cite{Varghese,IV_classic_review,Gamblin_Skehel,IVAreview}. 
As often in the virus realm, influenza is in fact a whole family of viruses
that have evolved slightly differently.
We will focus here on the two viruses where motility has been
evidenced experimentally \cite{Sakai_Saito_IVA,Sakai_Saito_IVC},
namely influenza-A (IVA) and influenza-C (IVC).

IVA has two spike proteins that interact with the host membrane: 
Hemaglutinin (HA) and Neuraminidase (NA), see Fig.~\ref{figsketch}. 
These are distinct $\sim10\,$nm sized entities (HA is a trimer, NA a tetramer)
that perform two distinct and 
mutually competing functions:
HA binds to sialic acid residues of glyco-peptides and lipids
coating the surface of our cells, while 
NA  acts antagonistically by hydrolytically cutting the
same 
sialic acid residue that HA binds to. 
Importantly, for steric reasons
the residue can be either
bound by one HA or by 
one NA molecule, but  not
by both at the same moment. 
In IVC, the two proteins are fused together into a single spike protein \cite{HEFref}, 
meaning the ``attaching spike'' and the ``cutting spike'' are co-localized.
But again, only one of the binding sites can interact with a glycan residue at a time.
Note that this renders IVA more flexible, as it can for instance polarize
its HA-NA distribution on its surface and engage in other mechanisms 
of motion than described here,
as discussed previously \cite{Fletcher}. This motion however is a much slower process, 
cf.~section \ref{sec_discussion}.
Due to lack of available experimental data, we study here the kinetics 
with IVA parameters and assume that the  values 
for IVC (once available) should be of similar order. 

The binding/unbinding kinetics of HA 
will be described via on/off rates $k_{on}$, $k_{off}$.
These have been characterized experimentally
\cite{Sauter1,Sauter2} yielding a dissociation constant
$K_d=\frac{k_{off}}{k_{on}}=1$-$5\,$(we use 2) mM, $k_{off}=10^{-1}$-$1\,(1)\,{\rm s}^{-1}$ and $k_{on}=0.01$-$1\,(0.5)\,{\rm mM}^{-1}{\rm s}^{-1}$.
NA transiently binds (with rates $k_1$, $k_{-1}$) and then enzymatically cuts 
the sialic residues with a rate $k_{cut}$, making the glycans irreversibly inactive for HA binding.
We hence use a Michaelis Menten description  
with a Michaelis constant $K_M = (k_{-1}+k_{cut}) / k_1 $. 
NA's enzymatic activity has also been measured \cite{Adams} 
to yield $K_M\simeq14.3$ mM and $k_{cut}\simeq15\,{\rm s}^{-1}$,
implying $V_{cut}=k_{cut}N_{NA}=15\,{\rm mM}{\rm s}^{-1}$
for a typical NA concentration of $N_{NA}=1$ mM.

\subsection{Contact interval}
\label{contact_interval}

During virus rolling all the force generation happens in the virus-substrate contact region, 
i.e.~the interface where the virus and the glycan-coated
substrate (cell membrane) meet. 
Here we roughly estimate the size of this region 
for a cross-section of a cylindrical (or spherical) virus
and will later substantiate the result 
by considering the full binding kinetics,
cf.~section \ref{detailed_balance}.

We assume that glycan chains are present at
a high concentration $G_{0}$ -- well in excess to spike proteins
(throughout this work we use $G_{0}=10\,{\rm mM}$ and for the HA spike concentration
$H_{0}=2\,{\rm mM}$, as estimated earlier \cite{virusPRL}).
The glycan chains 
are constantly binding to and unbinding from the HA spike proteins 
and elongate to a length  $l=R(1-\cos\phi)$ in that process,
where $\phi$ is the angle measured from the virus symmetry axis
and $R$ the virus radius.
If bound they gain a free energy
\begin{equation}
\Delta G = k_{B}T\ln\left(\frac{G_{0}}{K_{d}}\right)
\end{equation}
with $K_{d}$ the dissociation constant. 
In turn they have to pay the elastic energy of getting stretched
\begin{equation}
 E_{el}\left(\phi\right)=\frac{S}{2}l^{2}\simeq\frac{SR^{2}}{8}\phi^{4}\,\,\,{\rm for}\,\,\,\phi\ll1,
\end{equation}
where glycan chains were considered ideal linear
springs with spring constant $S\sim0.01$-$1\,k_{B}T/{\rm nm}^{2}$
 - a typical range for polymers of few nm length (we chose $0.1k_BT/nm^2$). 
Note that the cylindrical (or spherical) geometry of the virus results 
in a strong dependence on the angle $\phi$
and on the virus radius $R$ 
(which for IV is typically $\simeq50$ nm).
We neglect here the effect of NA binding for simplicity, since it is short lived compared to HA.

Balancing the
two energy terms yields the typical angular size of the contact zone with $\phi\in[-\phi_{c},\phi_{c}]$
to be
\begin{equation}\label{contact_angle}
 \phi_{c}=\left(\ln\left(\frac{G_{0}}{K_{d}}\right)\frac{8kT}{SR^{2}}\right)^{1/4}.
\end{equation}
In the following we assume the contact area size to be a constant,
even when the virus is rolling.

\subsection{Torque balance}
\label{sectorquebal}

The torque of an attached virus can be calculated using
the stretching force per linker $F_{el}=-\frac{\partial}{\partial l}E_{el}$
yielding
a torque $\propto SR^{2}\left(1-\cos\phi\right)\sin\phi\simeq\frac{1}{2}SR^{2}\phi^{3}$
for small angles.
If the linkers have an angular density $\rho_{HA}\,b\left(\phi\right)$
with $\rho_{HA}$ the angular density of HA spikes
and $b\left(\phi\right)$ the angular probability density of each linker being bound,
the total torque acting on the virus is just the integral 
over all bound linkers, 
\begin{equation}
m=-m_{0}\int_{-\phi_{c}}^{+\phi_{c}}b\left(\phi\right)\phi^{3}d\phi,\label{eq:Torque m}
\end{equation}
with $m_{0}=\frac{1}{2}SR^{2}\rho_{HA}$ the characteristic torque
scale. 

When the virus is rolling at typical angular speeds, experimentally 
$\omega\simeq1\,{\rm s}^{-1}$ \cite{Sakai_Saito_IVC},
using typical densities of linkers one can estimate all other torques,
e.g.~from hydrodynamics, to be negligible.
Therefore the torque balance $m=0$ has to hold (to very good approximation) at all times.

Note that for simplicity  we assumed here that there is no compression 
of the chains by the virus -- except at $\phi=0$ to fulfill force-balance. 
This simplification can be relaxed as explained 
in appendix \ref{compression}.

\subsection{Dynamics of bound linkers and free glycans}
\label{dynamics}

The binding of the HA spikes to the glycans can be described by a simple on-off kinetics.
Denoting the bound HA-glycan linker concentration by $B$, the 
unbound (open) HA concentration by $O$ and the free glycans by $G$,
one has $\partial_{t}B= k_{on}G\,O-k_{off}B$ and an equation
with opposite signs on the r.h.s.~for $\partial_{t}O$.
Obviously, as $O+B=H_{0}$ with $H_0$ the total number of HA,
one can immediately eliminate the equation for the open HA.

Adding the dynamics for the free glycans, one can write 
\begin{eqnarray}
\partial_{t}B+\omega\partial_{\phi}B & = & \,\,\,\,\,k_{on}G\left(H_{0}-B\right)-k_{off}B\,,\label{eq:ViroBoid B}\\
\partial_{t}G+\omega\partial_{\phi}G & = & -k_{on}G\left(H_{0}-B\right)+k_{off}B-f_{cut}\,,\,\,\quad\label{eq:ViroBoid G}
\end{eqnarray}
were we accounted for a (potential) rolling with angular velocity $\omega$, 
leading to advection of all concentration profiles as reflected by the second term on the l.h.s., 
and for the enzymatic cutting of $G$ by the NA spikes as reflected by the total cutting rate $f_{cut}$
acting as a sink. 

It is important to note that we made the approximation that the 
on/off-kinetics of HA-glycan binding   satisfies $\frac{k_{off}}{k_{on}}=K_{d}$. 
That is, we neglected that $K_d$ is in general stretching force- \cite{EvansRitchie}
and hence angle-dependent. This -- rather violent-looking  -- approximation 
allows for an analytical treatment. After having understood the general mechanism
of virus rolling, we show in section \ref{detailed_balance}
that the angle-dependence can be included -- in both continuum numerics and stochastic simulations --
and that this proper account of the detailed balance does not change the behavior qualitatively.
In other words, at this stage, in the simple model we use 
the stretch-dependence only to determine
the size of the contact interval, but not for the kinetics. 
Hence the glycan-binding profile of a static virus will be box-like 
(constant in the contact interval and zero outside)
while in reality it decays with $\exp(-\phi^4)$, cf.~section \ref{detailed_balance}.

The total cutting rate $f_{cut}$ can be approximated by a Michaelis-Menten kinetics,
as discussed in section \ref{spike_kinetics},
with a Michaelis constant $K_{M}$ 
and a cutting velocity of the sialic acid (glycan) residues, $V_{cut}=k_{cut}N_{NA}$ 
set by the enzymatic turnover rate $k_{cut}$ 
and the enzyme concentration $N_{NA}$, giving 
$ f_{cut}=\frac{V_{cut}G}{K_{\mathrm{M}}+G}$.

Finally, the minimal model reads 
\begin{eqnarray}
\partial_{t}B+\omega\partial_{\phi}B & = & k_{on}G\left(H_{0}-B\right)-k_{off}B,\,\label{eq:ViroBoidfinB}\\
\partial_{t}G+\omega\partial_{\phi}G & = & -k_{on}G\left(H_{0}-B\right)+k_{off}B
-\frac{V_{cut}G}{K_{\mathrm{M}}+G}\,\,\,\quad\label{eq:ViroBoidfinG}
\end{eqnarray}
and has to be solved on the contact interval $\phi\in[-\phi_{c},\phi_{c}]$ with 
$\phi_c$ given by Eq.~(\ref{contact_angle})
and
together with the torque balance constraint, i.e.~$m$ defined in 
Eq.~(\ref{eq:Torque m}) must be zero at all times.

\section{Steady rolling - scaling and numerics}
\label{steady}

Let us now discuss the just proposed -- deterministic and mean field-type  -- model
in the steady state on the scaling level. 
In steady state, the virus is either not moving at all or rolls with 
constant angular velocity $\omega$.
The first main question is: are states with $\omega\neq0$ possible
and if so to understand and analyze the causing mechanism.

\begin{figure}
\centering 
\includegraphics[width=0.4\textwidth]{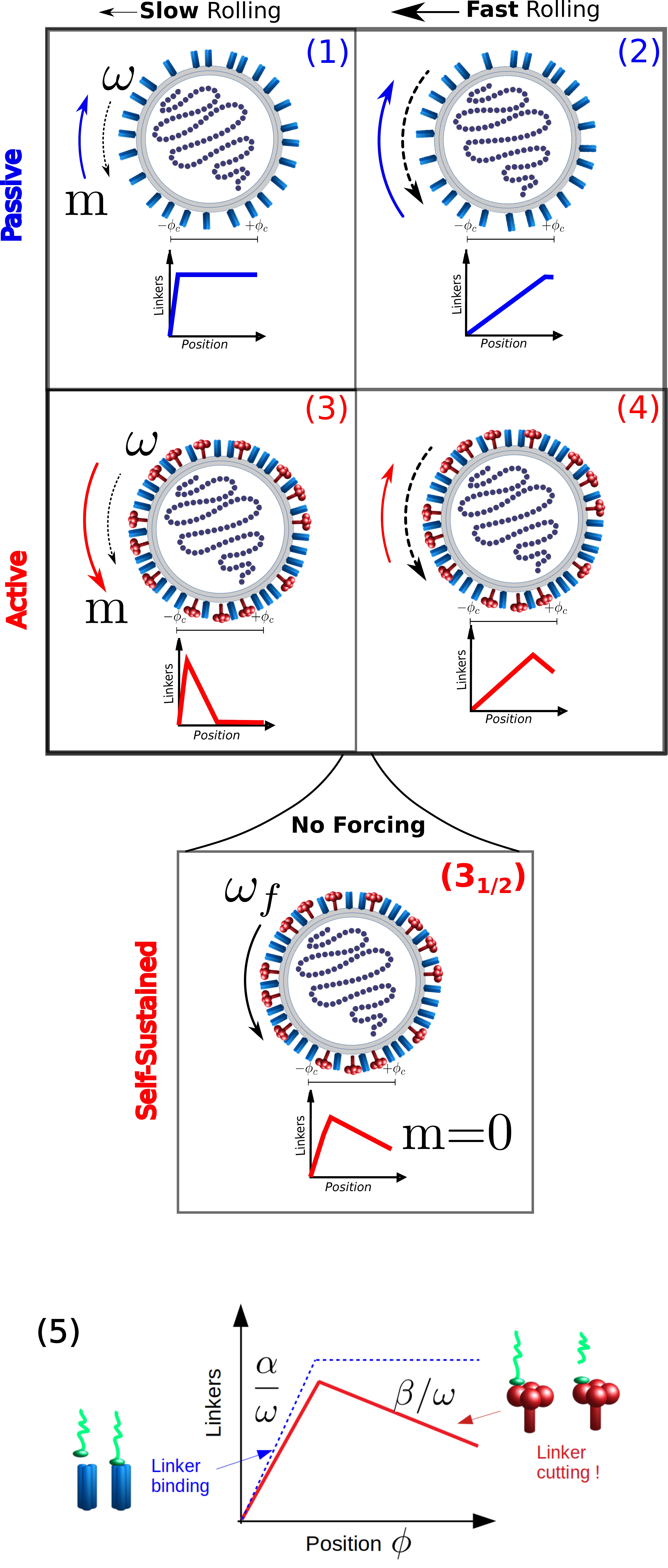}
\caption{Sketches of rolling virus cross-sections 
with representations of their bound linker distributions 
within the contact zone. Panels (1), (2): passive case (no NA activity)
for low vs.~high angular velocity $\omega$. 
Here the rolling must be due to an externally applied force
and the linker-induced torque, $m$, counteracts the rotation.
Panels (3), (4): active case with enzymatic NA activity, 
again for low and high $\omega$. The linker-induced torque accelerates (3) or decelerates (4)
the rotation. Panel ($3_{1/2}$): steady state, where the two regions
at the front and back counteract and the total torque is zero.
(5) Zoom of the passive (blue dashed) and active (red) profiles 
with characteristic profile slopes $\alpha$ and $\beta$ (see text)} 
\label{fig:VelRegimes}
\end{figure}

\subsection{Scaling arguments}
\label{sec_scale}

Before diving into the detailed calculations, let us try to explain
the rolling motion using simple scaling arguments.
Our focus lies on what the bound linker distribution looks like in the contact zone 
and the effects thereof for the torque balance.

First consider the simplest case of a virus that is forced to roll
by an externally applied torque -- as sketched in Fig.~\ref{figsketch}A) --
and that does not have enzymatic NA activity,
i.e.~only binds its HA to the glycans on the surface.
In rolling direction, the virus encounters unbound glycans and the HA needs time to bind,
which means the bound HA-glycan distribution in the co-moving virus frame
has the shape as sketched in
Fig.\ref{fig:VelRegimes}(1): it increases from zero with a certain slope
and levels off at a plateau value $B_{pl}$, corresponding 
to the mean bound linker distribution a non-moving virus would have.
The slope is determined by two quantities: 
first, the linker attachment rate 
\begin{equation}
\alpha=k_{on}H_0G_0 \label{alphadef}
\end{equation}
which is linked to the plateau value by $\alpha t_{pl}=B_{pl}$, 
with $t_{pl}$ the time it takes to establish
the equilibrium plateau.
The second is the actual rolling speed of the virus $\omega$.
Since $\omega t\sim\phi$ the slope in the angular distribution $B(\phi)$
is in fact given by $\frac{\alpha}{\omega}$.
Consequently, one expects the slope to be steep for a slowly rolling virus
and shallow if the virus rolls very fast, cf.~Fig.~\ref{fig:VelRegimes}(1) vs.~(2).

Now let us add a weak enzymatic cutting by NA.
In rolling direction, the binding kinetics will still dominate, 
for weak NA cutting rate $\omega$.
Hence the binding-induced positive slope in the attachment-region prevails.
For a steady rolling virus, however, with the distance from the front, 
the NA has linearly more time to cut off the glycans and hence one expects that the second, 
plateau region is transformed into an approximately linear, negative slope which we denote by $\beta$. 
This slope has in general a complicated parameter dependence, but it will be roughly proportional 
to the enzymatic cutting velocity 
\begin{equation}
\beta\propto V_{cut}. 
\end{equation}
By the same argument as before, the slope in the angular distribution $B(\phi)$
is given by $\frac{\beta}{\omega}$ and the slope at the back is 
steep for a slowly and shallower for a rapidly rolling virus,
cf.~Fig.~\ref{fig:VelRegimes}(3) vs.~(4).
Fig.~\ref{fig:VelRegimes}(5) shows a zoom on the contact zone
with the slopes labeled and with sketches of their different origins. 

What are the consequences of these linker distributions for the overall torque?
Note that a homogeneous, constant distribution $B(\phi)$ (and likewise any symmetric one)
is perfectly balanced.
Hence the linkers always missing at the front due to the time it needs to
establish the equilibrium distribution imply that there is an excess force acting on the back,
resulting in a torque counteracting the rotation, as indicated e.g.~in 
Fig.~\ref{fig:VelRegimes}(1) where the torque $m$ is acting against $\omega$.
Consequently, the attachment dynamics results in an effective friction.
In turn, the second, negative  slope implies the opposite: an excess force on the front,
accelerating the motion. Note that this is not forbidden thermodynamically:
the action of NA is an active process consuming, or rather cutting in an irreversible manner, the glycans.
Now it depends on the relative slopes/relative sizes of the two regions
whether the virus is overall accelerating, as in Fig.~\ref{fig:VelRegimes} (3), 
or decelerating as in Fig.~\ref{fig:VelRegimes} (4).
In between there is the possibility of a stationary state where the overal torque is zero,
as sketched in Fig.~\ref{fig:VelRegimes}($3_{1/2}$).
This exactly corresponds to the searched-for 
steady rolling state in the absence of external driving.

We can substantiate this argument by estimating the torque-angular velocity relation.
This relation is a {\it nonlinear} function -- and hence allows for non-trivial steady states -- 
due to the different $\omega$-dependencies of the two regions,
the attachment-dominated front and the cutting-dominated back.
In fact, for the attachment dynamics, if the respective region is small,
the torque evaluates to $m_{att}\propto -\frac{\alpha}{\omega} \cdot \phi_{pl}^2 \cdot \phi_c^3$,
where the first term is the profile's slope, the second term arises from the integral 
over the linear slope up to $\phi_{pl}$, the angle where the plateau begins, 
and $\phi_c^3$ is related to the geometric lever arm.
The minus sign indicates the direction, opposed to the rotation.
Since $\phi_{pl}=\omega t_{pl}$  
one gets $m_{att}\propto -\phi_c^3\,\alpha\,\omega$,  
i.e.~a ``Stokesian hydrodynamics''-like friction linear in $\omega$.
In contrast, for the cutting dynamics the contribution of the small front region does not matter
as cutting takes place everywhere at a uniform rate.
Hence the torque integral over the region of cutting  
is  $m_{cut}\propto \frac{\beta}{\omega} \cdot \phi_{c}^2 \cdot \phi_c^3$,
i.e.~depends on $\omega$ only via the slope.
The positive sign indicates 
its accelerating effect discussed above.
Clearly, the overall torque balance $0=m=m_{att}+m_{cut}$, 
allows for steady state solutions
of the type 
\begin{equation}
\omega\propto\pm\phi_c\sqrt{\frac{\beta}{\alpha}}\propto\sqrt{V_{cut}}.
\end{equation}
That is, the steady state rolling velocity grows rapidly for small cutting velocities $V_{cut}$.

\subsection{Time-angle correspondence and numerical solution}
\label{num_sec}

Let us  confirm numerically that self-rolling is possible.
In the steady state we can drop the time derivative in Eqs.~(\ref{eq:ViroBoidfinB}), (\ref{eq:ViroBoidfinG})
and have to solve 
\begin{eqnarray}
\omega\partial_{\phi}B & = & k_{on}G\,\left(H_{0}-B\right)-k_{off}B\,\label{eq:ViroBoidstatB}\\
\omega\partial_{\phi}G & = & -k_{on}G\,\left(H_{0}-B\right)+k_{off}B-\frac{V_{cut}G}{K_{\mathrm{M}}+G}\,,\,\,\quad\label{eq:ViroBoidstatG}
\end{eqnarray}
combined with
$b(\phi)=B(\phi)/H_{0}$ giving the closure condition in terms of the torque balance, Eq.~(\ref{eq:Torque m})  
\begin{equation}
m=-\frac{m_{0}}{H_{0}}\int_{-\phi_{c}}^{+\phi_{c}}B\left(\phi\right)\phi^{3}d\phi=0\,.\label{eq:Torquestatm}
\end{equation}
To do so, we can apply the following strategy, which could be called
``time-rotation angle correspondence''. The angular advection operator,
$\omega\partial_{\phi}$, and the time derivative operator, $\partial_{t}$,
can be treated on the same grounds by replacing time with the angle
scaled by the angular frequency: $t\rightarrow\frac{\phi+\phi_{c}}{\omega}$
(or $\phi=\omega t-\phi_{c}$). The time window corresponding to passing
the contact interval $[-\phi_{c},\phi_{c}]$ is then $[0,T]$ with
$T=\frac{2\phi_{c}}{\omega}$.

This observation suggests to think of the steady state as a dynamic
relaxation of the concentrations on the time interval $[0,T]$. One
can hence treat the problem as an initial value problem at one
boundary of the interval (the one in rolling direction) with $B\left(0\right)=0$
and $G\left(0\right)=G_{0}$ (in the not yet visited region, nothing
has bound and no glycan has been consumed yet): the solution of the
dynamic problem $B_{dyn}\left(t\right)$ on a time interval $[0,T_{test}]$
can be obtained without any reference to the actual angular frequency
$\omega$.

In a second step $B_{dyn}\left(t\right)$ 
can be used to obtain 
the torque by evaluating 
\begin{equation}
m=-\frac{m_{0}}{H_{0}}\int_{0}^{t_{up}=2\phi_{c}/\omega}B_{dyn}\left(t\right)\left(\omega t-\phi_{c}\right)^{3}\omega dt\label{torque_lin_num}
\end{equation}
Note that high frequencies correspond to taking the integral over
a short time interval and vice versa. Given the curve $B_{dyn}\left(t\right)$
on a large enough interval, $[0,T_{test}]$ with $T_{test}>T$ for all
considered $\omega$ values, one can now scan the upper boundary $t_{up}=2\phi_{c}/\omega$
by varying the angular frequency
and hence determine $\omega$ such that $m\left(\omega\right)=0$,
which is a simple root finding problem. 
Solving along the same lines for  $m\left(\omega\right)=m_{ext}$,
with an external torque $m_{ext}$, allows also to obtain the torque-angular velocity relation.

\begin{figure}
\includegraphics[width=0.5\textwidth]{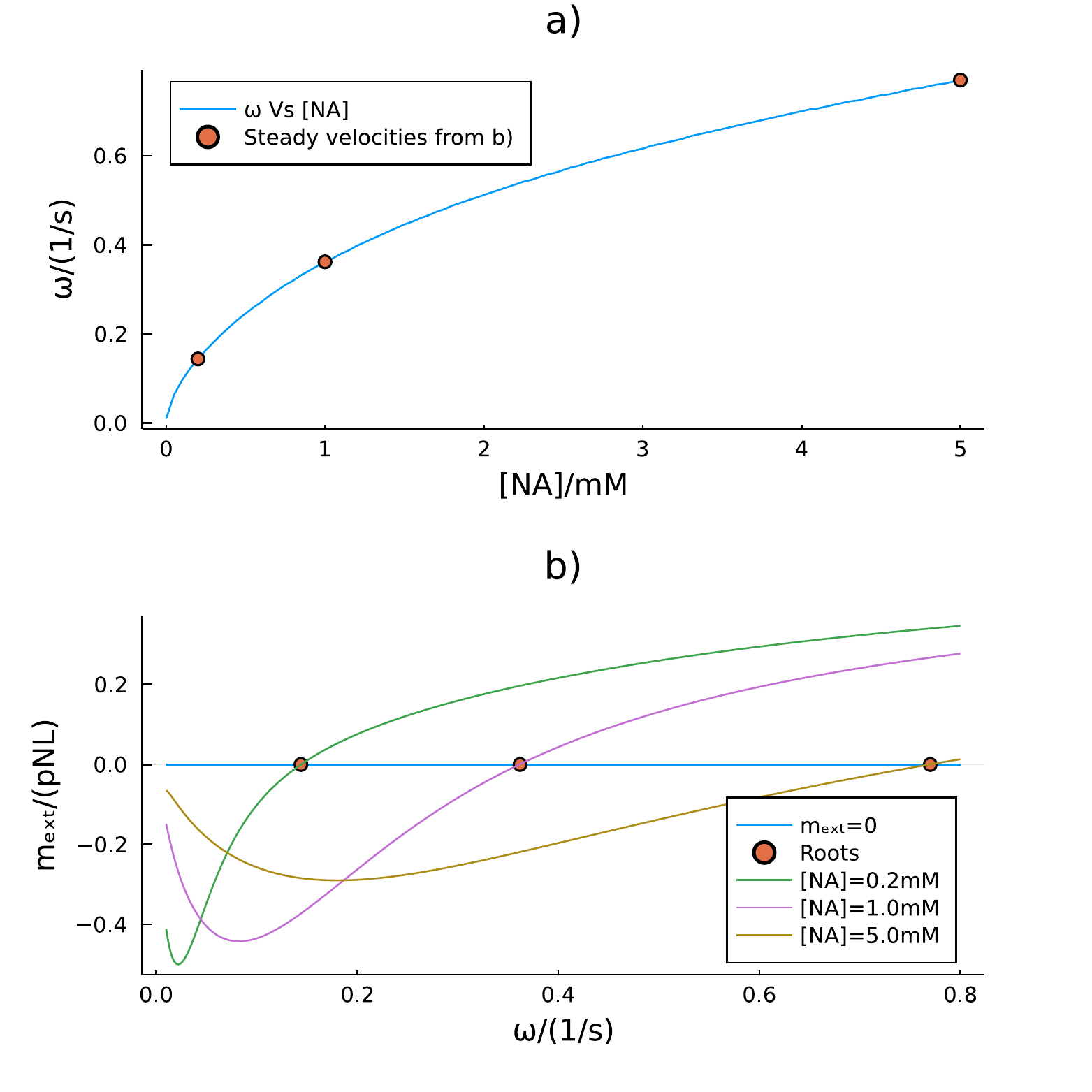}
\caption{\label{numresults}
a) Angular velocity $\omega$ of the steady rolling state as a function 
of NA concentration $N_{NA}=[NA]$.
One abtains a square-root behavior, 
$\omega\propto\sqrt{N_{NA}}$, as suggested by scaling.
b) Adding an external torque $m_{ext}$ allows to obtain the torque-angular velocity relation,
which is a nonlinear function of $\omega$.
Three different values of NA concentration are shown,
the respective free rolling velocities (for $m_{ext}=0$) are also indicated in a).
}
\end{figure}

Fig.~\ref{numresults} a) shows the angular velocity of the steady rolling state, 
obtained numerically, as a function of enzymatic activity, i.e.~NA concentration.
One clearly sees the square-root behavior as obtained by scaling,
$\omega\propto\sqrt{\beta}$ with $\beta\propto V_{cut}\propto N_{NA}$. 
Fig.~\ref{numresults} b) shows the torque-angular velocity relation obtained numerically
for three values of NA concentration. 
Free rolling corresponds to $m_{ext}=0$. 

The results obtained numerically and shown in Fig.~\ref{numresults}
are intriguing: in fact, any -- even a small -- enzyme activity leads to finite rolling motion
in this mean field-type model. The response of the motor close to free
rolling is as expected -- an assisting torque speeds up the rolling and a counter torque 
slows it down, but there is a strong nonlinear dependence for higher counter torques. 
Both these findings can be understood analytically and can be related to the
parameters of the spike dynamics, as shown in the next section.
They will then be critically compared to a model including stochastic fluctuations
in section \ref{stochastic}.

\section{Steady rolling - Analytical theory}
\label{steadyanalyt}

In the steady state 
much analytical insight can be gained using 
the main ideas introduced in the last chapter, i.e.~the time-angle-correspondance 
and the "line-approximations" of the linker profiles, as sketched in Fig.~\ref{fig:VelRegimes}. 

\subsection{Analytical solution for forced rolling without enzymatic activity}
\label{analytic_forced_roll}

We first consider the passive case,
i.e.~in absence of catalytic activity. 
That is, we assume that the virus is forced to roll with a given
steady-state angular velocity $\omega$ by a weak, externally applied
torque. 
For $V_{cut}=0$ and in the steady state, Eqs.~(\ref{eq:ViroBoidstatB},
\ref{eq:ViroBoidstatG}) imply the conservation law $\partial_{\phi}\left(B+G\right)=0$.
Assuming a homogeneous initial glycan coverage $G_{0}$, we can hence
rewrite $G(\phi)=G_{0}-B(\phi)$. This reduces the problem to a single
equation 
\begin{equation}
\omega\partial_{\phi}B=k_{on}\left(G_{0}-B\right)\,\left(H_{0}-B\right)-k_{off}B
\end{equation}
which, for the initial condition $B(-\phi_{c})=0$ (rolling to the
left), can be solved exactly 
\begin{equation}
B\left(\phi\right)=\frac{C_{0}-C_{1}}{2}-\frac{C_{1}}{\frac{C_{0}+C_{1}}{C_{0}-C_{1}}e^{\frac{C_{1}k_{on}}{\omega}\left(\phi+\phi_{c}\right)}-1}.\label{eq:B-equil}
\end{equation}
Here $C_{0}=H_{0}+G_{0}+K_{d}$, $C_{1}=\sqrt{C_{0}^{2}-4H_{0}G_{0}}$
are constants determined by the concentrations and reaction kinetics.
The solution can also be given in time-domain: 
\begin{equation}
B\left(t\right)=\frac{C_{0}-C_{1}}{2}-\frac{C_{1}}{\frac{C_{0}+C_{1}}{C_{0}-C_{1}}e^{C_{1}k_{on}t}-1}.\label{eq:B-equil_time}
\end{equation}

\begin{figure}
\includegraphics[width=0.35\textwidth]{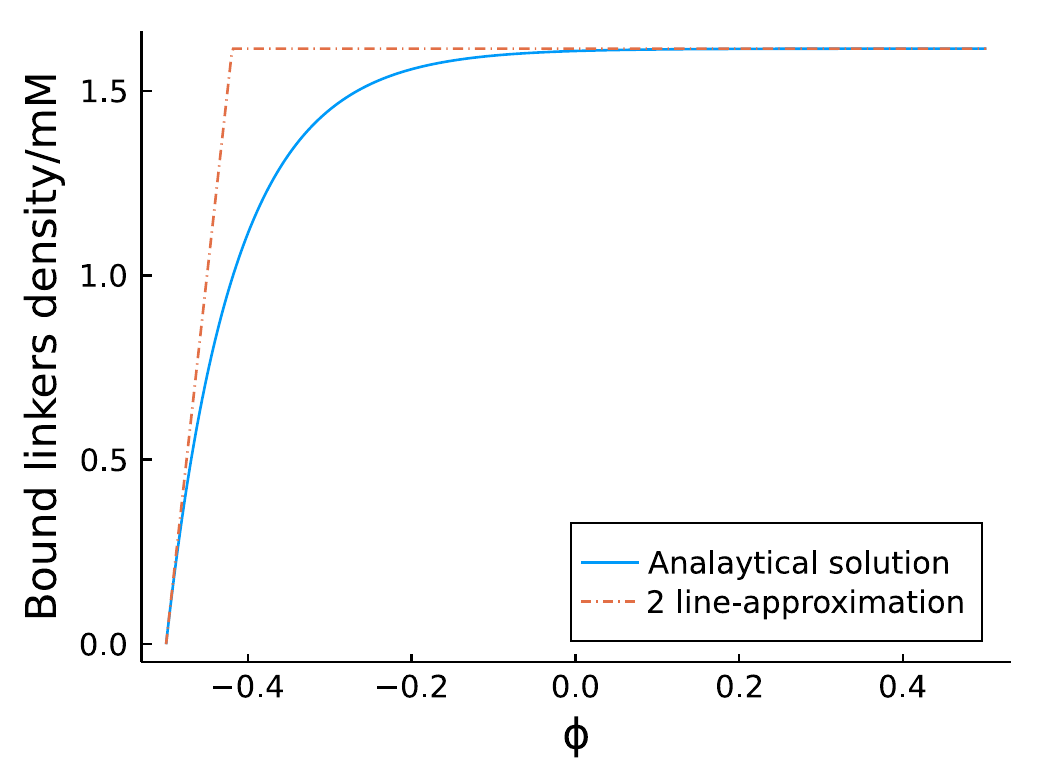}
\caption{\label{Bpassive}
The blue curve shows the bound linker profile 
on the contact angle interval $[-\phi_c,\phi_c]$ as given by Eq.~(\ref{eq:B-equil}). 
The virus is forced to roll to the left.
The fact that the linkers need time to bind to the newly encountered substrate
is reflected by the increase on the left leveling to a plateau, cf.~Fig.~\ref{fig:VelRegimes} 
and the discussion is section \ref{sec_scale}.
The dashed lines show the two-line approximation given by 
Eq.~(\ref{eq:B-2line-approx-3angle}).
Parameters as given in section \ref{basic}, implying $\phi_c\simeq0.5$.
}
\end{figure}

The resulting bound linker profile  on the contact angle interval $[-\phi_c,\phi_c]$
is shown in Fig.~\ref{Bpassive}, cf.~also the cases sketched as (1) and (2) in Fig.~\ref{fig:VelRegimes}. It is characterized by an increase of the 
bound HA-glycan links leveling at a plateau value of 
\begin{equation}
B_{pl}=\frac{C_{0}-C_{1}}{2}.
\end{equation}
For simplicity, and to be able to proceed with a perturbative approach taking 
enzymatic activity into account, we approximate the exact profile
by two lines: first, in the region of its rapid increase, $B$ is
approximated by the slope at the front, and in the second region by
its plateau value. This is most transparent in time-space where one has 
\begin{equation}
B\left(t\right)=\begin{cases}
\alpha t & \text{ for \ensuremath{0\le t\le t_{m}}}\\
\alpha t_{m} & \text{ for \ensuremath{t_{m}\le t\le T}}
\end{cases}\label{eq:B-2line-approx-time}
\end{equation}
for $t\in\left[0,T\right]$ with $T=\frac{2\phi_{c}}{\omega}$ as
before and 
\begin{equation}
\alpha=k_{on}H_{0}G_{0}\,,\,\,\,t_{m}=
\frac{B_{pl}}{\alpha}\label{alpha_tm}
\end{equation}
with $\alpha$ the initial slope (i.e.~the linker
binding velocity) and $t_{m}$ the time needed to reach the plateau/maximum
(cf.~$t_{pl}$ discussed in section \ref{sec_scale}). 
In angle-space one has 
\begin{equation}
B\left(\phi\right)=\begin{cases}
\frac{\alpha}{\omega}(\phi+\phi_{c}) & \text{ for \ensuremath{\phi\in[-\phi_{c},\phi_{pl}]}}\\
\frac{\alpha}{\omega}(\phi_{pl}+\phi_{c})=B_{pl} & \text{ for \ensuremath{\phi\in[\phi_{pl},\phi_{c}]}}
\end{cases}\label{eq:B-2line-approx-3angle}
\end{equation}
with $\phi_{pl}$ the angle where
the plateau is reached. 
This angle space view is especially transparent to
derive the scaling discussed earlier, cf.~appendix 
\ref{angle_space_scaling}.

\subsection{Multiple-line approximations}
\label{multiline}

One can proceed the analysis by using the approximations of the bound linker
profiles by several lines. Typically, two lines are needed, cf.~the cases 
sketched in Fig.~\ref{fig:VelRegimes} as (1),(2),(4). 
When the linkers are completely cut at the back, 
three lines are needed,  cf.~Fig.~\ref{fig:VelRegimes} (3). 
These multiple-line approximations also allow us to treat the effect of enzymatic
activity analytically and to discuss the generic physics of the torque-angular
velocity relation associated with a given bound linker profile.

\subsubsection{Case of no enzymatic activity}

This is the case just discussed 
where $B(t)$ is approximately given by Eq.~(\ref{eq:B-2line-approx-time}). 
The clear advantage of a line approximation 
is that we can calculate the torque integral analytically. From Eq.~(\ref{eq:Torque m})
we have to evaluate 
\begin{equation}
m=-\frac{m_{0}\omega^{4}}{H_{0}}\int_{0}^{T}\hspace{-1mm}\left(t-\frac{T}{2}\right)^{3}B\left(t\right)dt
\end{equation}
with $T=\frac{2\phi_c}{\omega}$.
Importantlly, we can write this as an integral of the constant plateau
value over the full range, $\propto\int_{0}^{T}\hspace{-1mm}\left(t-\frac{T}{2}\right)^{3}\hspace{-1mm}B_{pl}\,dt=0$,
which vanishes due to symmetry, plus an integral over the first zone
only (the one at the front in rolling direction) 
\begin{equation}
m=-\frac{m_{0}\omega^{4}}{H_{0}}\int_{0}^{t_{m}}\hspace{-1mm}\left(t-\frac{T}{2}\right)^{3}\hspace{-1mm}\alpha(t-t_{m})dt\,.
\label{fronteq}
\end{equation}
This highlights the fact that the torque arises from the imbalance/asymmetry
of the linker distribution associated with linkers having to form
in rolling direction in the region $0\le t\le t_{m}$.

Before solving in full generality, we can assume that the plateau
is rapidly reached, i.e.~$t_{m}\ll T$, yielding 
\begin{eqnarray}
 m&\simeq& \frac{m_0\omega^4}{H_0} \alpha \left(\frac{T}{2}\right)^{\hspace{-1mm}3}\hspace{-1mm} \int_{0}^{t_m} (t-t_m) dt
=
-\frac{m_0}{H_0}\frac{\phi_c^3}{2}\frac{B_{pl}^2}{\alpha}\omega,\quad\,\,\,
\label{pass_fric}
\end{eqnarray}
where we used $T=\frac{2\phi_c}{\omega}$ and 
$t_m=B_{pl}/\alpha$.
Eq.~(\ref{fronteq}) can also be integrated completely to yield 
\begin{equation}
m=\frac{m_{0}}{H_{0}}\frac{\alpha}{80}\omega^{4}T^{5}p\left(\frac{t_{m}}{T}\right)
\label{torque_noact}
\end{equation}
with the polynomial 
\begin{equation}
p\left(x\right)=-x^{2}\left(5-10\,x+10\,x^{2}-4\,x^{3}\right)\label{polynomial}
\end{equation}
for $t_m\leq T$;  for $t_m>T$ one has to use $p(1)=-1$.

Let us discuss the limits. 
For very slow rolling, cf.~Fig.~\ref{fig:VelRegimes}(1), 
the first, attachment-dominated regime shrinks with only the plateau left, 
and the torque hence vanishes due to symmetry. 
For $t_{m}\ll T$, corresponding to intermediate rolling speeds and
sufficiently rapid build-up of the plateau, one has 
$p(x)\simeq-5x^{2}$ which agrees with Eq.~(\ref{pass_fric}). 
Hence for intermediate rolling speed one gets for the torque-velocity relation for passive rolling, 
\begin{equation}
\label{mdiss}
m_{diss}\left(\omega\right)=-\xi_{diss}\omega\,\,,\,\,\,\,\xi_{diss}=\frac{m_{0}}{H_{0}}\frac{\phi_{c}^{3}}{2}\frac{B_{pl}^{2}}{\alpha}.
\end{equation}
This is a purely frictional torque, acting against the motion
and linear in $\omega$.
The effective friction constant, $\xi_{diss}$, is determined by both the slope, $\alpha$, 
and the plateau value, $B_{pl}$, 
of the bound linker distribution, 
as well as the size of the contact interval $\phi_c$, which themselves
contain all system parameters.

For faster rolling, the friction is determined by $p(x)$.
For very fast rolling, faster even than in Fig.~\ref{fig:VelRegimes}(2), 
the linking is so slow compared 
to the rolling that there will be no plateau at all.
In that case, one evaluates $m=-\frac{m_{0}}{H_{0}}\alpha\omega^{4}T^{5}\frac{1}{80}$,
which corresponds to the value $p(1)$. The torque is hence still frictional
and due to $T=\frac{2\phi_{c}}{\omega}$ scales with $\omega^{-1}$.
Hence in this regime, the faster the rolling the less friction: less linkers bind because of limited time 
and hence fewer resist the motion.
Overall, friction first increases, then is given by $p(x)$ and then decreases again.

\subsubsection{Case of weak enzymatic activity}

In the presence of enzymatic NA activity, cf.~Fig.~\ref{fig:VelRegimes}(3)-(5),
the first region of increasing linker density is still present (in fact, only slightly modified), 
while the plateau is transformed into a slowly decreasing function, 
implying $B(t)$ having a maximum around $t_{m}$.
Focusing on cases (4),(5) where the glycan is not completely cut at the back,
we can use the following parameterization 
\begin{equation}
B\left(t\right)=\begin{cases}
\alpha t & \text{ for \ensuremath{0\le t\le t_{m}}}\\
\alpha t_{m}-\beta\left(t-t_{m}\right) & \text{ for \ensuremath{t_{m}\le t\le T}}
\end{cases}\label{eq:B-2line-approx}
\end{equation}
with two non-negative constants $\alpha>0,\beta\ge0$ 
(with $\beta=0$ corresponding to the case without NA activity just discussed, 
without maximum but a plateau).
At the moment $\beta$ is just a parameter. 
In section \ref{perturb} we will determine it 
as a function of the underlying model parameters using perturbation theory.

We can again bring the torque integral in a convenient form (using
that the integral over the plateau over the full contact area vanishes)
and get 
\begin{eqnarray}
m & = & -\frac{m_{0}\omega^{4}}{H_{0}}\left[\int_{0}^{t_{m}}\hspace{-1mm}\alpha(t-t_{m})\left(t-\frac{T}{2}\right)^{3}\hspace{-1mm}dt\right.\nonumber \\
 &  & \left.\quad\qquad-\int_{t_{m}}^{T}\hspace{-1mm}\beta(t-t_{m})\left(t-\frac{T}{2}\right)^{3}\hspace{-1mm}dt\right]\label{2Lwithbeta}
\end{eqnarray}
If the plateau is rapidly reached, $t_{m}\ll T$,
one obtains
\begin{eqnarray}
m
=-\frac{m_{0}}{H_{0}}\left[\alpha\omega\frac{1}{2}\left(t_{m}\right)^{2}\phi_{c}^{3}-\frac{\beta}{\omega}\frac{2}{5}\phi_{c}^{5}\right].\label{forcevelscalemain}
\end{eqnarray}
(this is also derived in Eq.~(\ref{forcevelscale})).
Importantly, the second term has the opposite sign -- it is active
-- since now $t>t_{m}$ in the integration. Secondly, the second
term is independent of $t_{m}$ for small $t_{m}$, which leads to
the $1/\omega$ dependence. 

Eq.~(\ref{2Lwithbeta}) can again be integrated completely, yielding
\begin{equation}
m=\frac{m_{0}\omega^{4}}{H_{0}}T^{5}\left\{ \frac{\alpha}{80}\,p\left(\frac{t_{m}}{T}\right)
+\frac{\beta}{80}\left[p\left(\frac{t_{m}}{T}\right)+1\right]\right\}. 
\end{equation}
We hence get the relation 
\begin{equation}
\frac{80H_{0}}{\left(\alpha+\beta\right)\omega^{4}T^{5}}\frac{m}{m_{0}}=p\left(\frac{t_{m}}{T}\right)+\frac{\beta}{\alpha+\beta},
\end{equation}
from which one can draw general conclusions:
First, $\omega=0$, implying $T=\infty$ and hence $\frac{t_{m}}{T}=0$ and
$p\left(\frac{t_{m}}{T}\right)=0$, is a solution for $\beta=0$. This
corresponds to the static, non-rolling case without activity. Next,
the polynomial $p\left(\frac{t_{m}}{T}\right)$ 
is always negative (except for $\frac{t_{m}}{T}=0$), with its modulus
increasing with $t_{m}$. 
Therefore, only when the term $\beta>0$ is present, there is the possibility
to ensure $m\propto p\left(\frac{t_{m}}{T}\right)+\frac{\beta}{\alpha+\beta}=0$,
i.e.~torque balance, for finite $\frac{t_{m}}{T}$ and hence $\omega$. On the other hand, for
any $\beta>0$ there is in fact always a solution $t_{m}=t_{m}^{*}$
at which the torque vanishes, namely 
\[
p\left(\frac{t_{m}^{*}}{T}\right)=-\frac{\beta}{\alpha+\beta}.
\]
We have hence shown that, at least in two-slope approximation, \textit{any
finite enzyme activity will induce motion} in the simple model. 

\subsubsection{Case of slow rolling and high enzymatic activity: three-line approximation}

Let us discuss the case sketched in Fig.~\ref{fig:VelRegimes}(3).
So far we had assumed that the bound linkers cannot decay to zero in the time interval
$[0,T]$.
However, for small $\omega$ and high $\beta$, 
the two-line approximation breaks
down if the bound linker concentration becomes negative in $[t_{0},T]$
with $\alpha t_{m}-\beta\left(t_{0}-t_{m}\right)=0$ or 
\begin{equation}
t_{0}=\frac{\alpha+\beta}{\beta}t_{m}.
\end{equation}
To cover the whole range of frequencies, 
we generalize the profile to a Three-Line-Approximation (3LA) by writing
\begin{equation}
B\left(t\right)=\begin{cases}
\alpha t & \text{ for \ensuremath{0\le t\le t_{m}}}\\
\alpha t_{m}-\beta\left(t-t_{m}\right) & \text{ for \ensuremath{t_{m}\le t\le t_{0}}}\\
0 & \text{ for \ensuremath{t_{0}\le t\le T}}
\end{cases}\label{eq:B-3line-approx}
\end{equation}
We simplify the evaluation of the torque integral as follows: in the
limit of small $\omega$ (i.e.~large $T$) and not too high $\beta$
(such that $t_{0}>t_{m}$ and $\frac{\alpha+\beta}{\beta}\simeq\frac{\alpha}{\beta}$)
the integral over the second region dominates over the first, which
is very small (rapid rise to the plateau), and the third
region does not contribute anyways. It is hence enough to evaluate
\[
m\simeq-\frac{m_{0}\omega^{4}}{H_{0}}\int_{t_{m}}^{t_{0}}\left(t-\frac{T}{2}\right)^{3}\left(\alpha t_{m}-\beta\left(t-t_{m}\right)\right)\,dt
\]
Moving the lower boundary to $0$ and using $t_{0}\simeq\frac{\alpha}{\beta}t_m$
(as discussed above) one gets by neglecting terms which are a factor
$t_{m}/t_{0}$ smaller
\begin{equation}\label{m3LA}
m\simeq-\frac{m_{0}\omega^{4}}{H_{0}}\frac{\beta}{80}T^{5}p\left(\frac{t_{0}}{T}\right)
\end{equation}
with still the same polynomial, cf.~Eq.~(\ref{polynomial}).

\subsubsection{Full torque-angular velocity relation}
\label{full_fv}

Using $T=\frac{2\phi_{c}}{\omega}$ one can introduce the characteristic
frequency 
\begin{equation}
\omega_{0}=\frac{2\phi_{c}}{t_{0}}=\frac{2\phi_{c}}{t_{m}}\frac{\beta}{\alpha}
\end{equation}
to replace the argument of the polynomial in Eq.~(\ref{m3LA}) by $\omega/\omega_{0}$.
Rescaling as $\tilde{\omega}=\omega/\omega_{0}$ one gets
the following torque-angular velocity relation for small frequencies, $\tilde{\omega}<1$, i.e. $\omega<\omega_0$: 
\begin{equation}\label{tvr_smallom}
m=m_{0}\frac{B_{pl}}{H_{0}}\frac{\phi_{c}^{4}}{5}\tilde{\omega}\cdot
\left(-4\,\tilde{\omega}^{3}+10\,\tilde{\omega}^{2}-10\,\tilde{\omega}+5\right).
\end{equation}

For larger frequencies , $\tilde{\omega}>1$, we can use  Eq.~(\ref{forcevelscalemain}),
which can be rescaled the same way to yield
\begin{equation}\label{tvr_highom}
m=-m_{0}\frac{B_{pl}}{H_{0}}\frac{\phi_{c}^{4}}{5}\left(5\frac{\beta}{\alpha}\tilde{\omega}
-\frac{1}{\tilde{\omega}}\right).
\end{equation}

We hence have established the full torque-angular velocity relation
for a rolling virus. While $\alpha$ is determined simply by the HA-glycan on-kinetics,
$\beta$ is yet not specified. We will determine it in the next section 
by a perturbation expansion of the non-enzymatic state and postpone 
the discussion of the steady rolling and the torque-angular velocity relation to section
\ref{ana_discussion}.

\subsection{Including enzymatic activity: perturbative solution }
\label{perturb}

We now include the enzymatic activity of NA and
treat it as a perturbation of the forced-rolling steady state, obtained 
in section \ref{analytic_forced_roll}, 
where HA-glycan binding leads to the two-line profile 
with a rapid increase at the front (characterized by $\alpha$) 
leveling off to a plateau.

We hence assume that $\epsilon=V_{cut}/\alpha$ is a small parameter i.e.~the
enzyme cutting activity is small against the binding kinetics. 
We then can write 
\begin{equation}
B=B_{1}+\epsilon B_{2}+\ldots\,,\,\,G=G_{1}+\epsilon G_{2}+\ldots.
\end{equation}
such that the zero order $\mathcal{O}(\epsilon^{0})$ is just the passive (i.e.~forced rolling)
case of section \ref{analytic_forced_roll}. 

To next order $\mathcal{O}(\epsilon)$ one has to consider 
\begin{eqnarray}
\omega B_{2}' & = & k_{on}H_{0}G_{2}-k_{off}B_{2}-k_{on}\left(G_{1}B_{2}+B_{1}G_{2}\right),\nonumber \\
\omega G_{2}' & = & -k_{on}H_{0}G_{2}+k_{off}B_{2}+k_{on}\left(G_{1}B_{2}+B_{1}G_{2}\right)\nonumber \\
 &  & -\frac{k_{on}H_{0}G_{0}G_{1}}{K_{M}+G_{1}}.\,\,\,\,\,\,
\end{eqnarray}
As the binding kinetics is assumed to be fast, 
we use an adiabatic approximation, $\omega B_{2}'\simeq0$. 
Then the equations for $G_{2}$ and $B_{2}$ simplify to 
\begin{equation}
\omega G_{2}'=-\frac{k_{on}H_{0}G_{0}G_{1}}{K_{M}+G_{1}}\,,\,\,\,
B_{2}
=\frac{H_{0}-B_{1}}{K_{d}+G_{1}}G_{2}\,.
\end{equation}
Neglecting the boundary layer close to the front, we can replace $B_{1}\simeq B_{pl}$,
$G_{1}\simeq G_{pl}=G_0-B_{pl}$ by their plateau values. 
Also note that $G_{2}(t=0)=0$, since the boundary conditions  at the front are already
fulfilled in zero order. 
Transforming to time space we
get the dynamical profile for the glycans  
$G_{2}\left(t\right)=-\frac{k_{on}H_{0}G_{0}G_{pl}}{K_{M}+G_{pl}}t$
and hence for the bound linkers 
\begin{equation}
B_{2}\left(t\right)=-\frac{H_{0}-B_{pl}}{K_{d}+G_{pl}}\frac{k_{on}H_{0}G_{0}G_{pl}}{K_{M}+G_{pl}}t\,.
\end{equation}
Transforming back to angle space and using  $\epsilon=V_{cut}/\alpha$ 
where $\alpha=k_{on}H_0G_0$, yields \cite{comment1}
$B_{2}=-\alpha f\,\frac{\phi+\phi_c}{\omega}$
or
\begin{equation}
\beta=V_{cut}f\,,\,\,\,f= \frac{H_0-B_{pl}}{K_d+G_{pl}} \frac{  G_{pl}}{K_M+G_{pl}}.
\end{equation}

This result directly implies an expansion for the torque,
\begin{equation}
m=m_{1}+\epsilon m_{2},
\end{equation}
where $m_{1}=m_{diss}$ is the passive torque, given in Eq.~(\ref{mdiss}), 
and $\epsilon m_{2}=m_{act}$ is the active torque. 
Explicitly one obtains
\begin{equation}
m_{act}=\frac{p_{act}}{\omega}\,,\,\,p_{act}=\frac{m_{0}}{H_{0}}f\,\frac{2\phi_{c}^{5}}{5}V_{cut}\,.\label{mactfact}
\end{equation}
Here $p_{act}$ is the power injected by NA operation. The active
torque is positive (since $B_{pl}<H_{0}$), it is proportional to
$V_{cut}$ and has a $1/\omega$ dependence, unlike the passive one
which is linear in $\omega$.

\subsection{Discussion of rolling velocity and force-velocity relation}
\label{ana_discussion}

Let us now discuss the results obtained analytically.
In the case of weak enzymatic activity, we can combine 
Eq.~(\ref{mdiss}) for the dissipative torque and Eq.~(\ref{mactfact})
for the active driving torque
to the torque balance $m_{diss}+m_{act}=0=-\xi_{diss}\omega+p_{act}/\omega$.
This immediately implies a pitchfork bifurcation for the steady-state rolling velocity
\begin{equation}
\omega=\pm\sqrt{\frac{p_{act}}{\xi_{diss}}}\propto\phi_{c}\sqrt{f}\,\frac{\sqrt{\alpha V_{cut}}}{B_{pl}}\,.\label{om_virus}
\end{equation}
This is exactly what had been observed in Fig.~\ref{numresults}a),
the velocity scaling like $\omega\propto\pm\sqrt{\beta}$ 
with $\beta\propto V_{cut}$ with $V_{cut}$ linear in the 
NA concentration. 
As the total torque is zero, the torque scale $m_{0}=\frac{1}{2}SR^{2}\rho_{HA}$ 
cancels out. Nevertheless the parameters $S$ (linker stiffness) and $R$ (radius of the virus)
are still present since they enter the contact interval size $\phi_{c}$. 
The result in fact depends on all model parameters,
especially the kinetics of attachment/detachment and cutting contained in $f$.

Using the specific parameter values for influenza given in section \ref{spike_kinetics},
Eq.~(\ref{om_virus}) yields values of the order of $\omega=0.4\,{\rm s}^{-1}$. 
This compares well to the experimentally measured values by
Sakai et al.~\cite{Sakai_Saito_IVA,Sakai_Saito_IVC}:
there, translational speeds of of $v\simeq10-30\,{\rm nm/s}$ were reported,
corresponding to $\omega$ between $0.2-0.6\,{\rm s}^{-1}$
for physiological NA activity.

To critically discuss the torque-angular velocity relation,
we will introduce
\begin{equation}
A=\frac{\beta}{\alpha},
\end{equation}
which is an ``activity parameter'' proportional to the enzymatic activity. 
Adding now a term $5A\tilde{\omega}$ to Eq.~(\ref{tvr_smallom}), 
which represents a small correction since $\beta\ll\alpha$, 
allows us to 
combine the two obtained limits, Eq.~(\ref{tvr_smallom}) and
Eq.~(\ref{tvr_highom}) to one continuous curve,
\begin{equation}
\tilde{m}\left(\tilde{\omega}\right)\approx-\begin{cases}
\tilde{\omega}\cdot\left(-4\tilde{\omega}^{3}+10\tilde{\omega}^{2}-10\tilde{\omega}+5\left(1-A\right)\right) & 
\hspace{-2mm}\text{; \ensuremath{\tilde{\omega}}\ensuremath{\ensuremath{\le}1}}\\
\tilde{\omega}^{-1}-5A\tilde{\omega} & 
\hspace{-2mm}\text{; \ensuremath{\tilde{\omega}}\ensuremath{>1}}
\end{cases}\label{eq:Motor-Curve}
\end{equation}
where in addition we non-dimensionalized the torque $\tilde{m}=\frac{m}{m_{c}}$
with $m_c=m_{0}\frac{B_{pl}}{H_{0}}\frac{\phi_{c}^{4}}{5}$.
Note that, maybe counterintuitively, the activity is now in front of
the passive torque for the branch $\tilde{\omega}>1$. 
This is due to the characteristic frequency, 
$\omega_{0}=\frac{2\phi_{c}}{t_{0}}=\frac{2\phi_{c}}{t_{m}}A$,
being proportional to the activity parameter:
the higher the activity, the larger the rolling frequency must be
to prevent the depletion zone at the back (the $B=0$ region) to occur, 
and to stay in the regime $\tilde{\omega}>1$.  

\begin{figure}
\centering \includegraphics[width=1\linewidth]{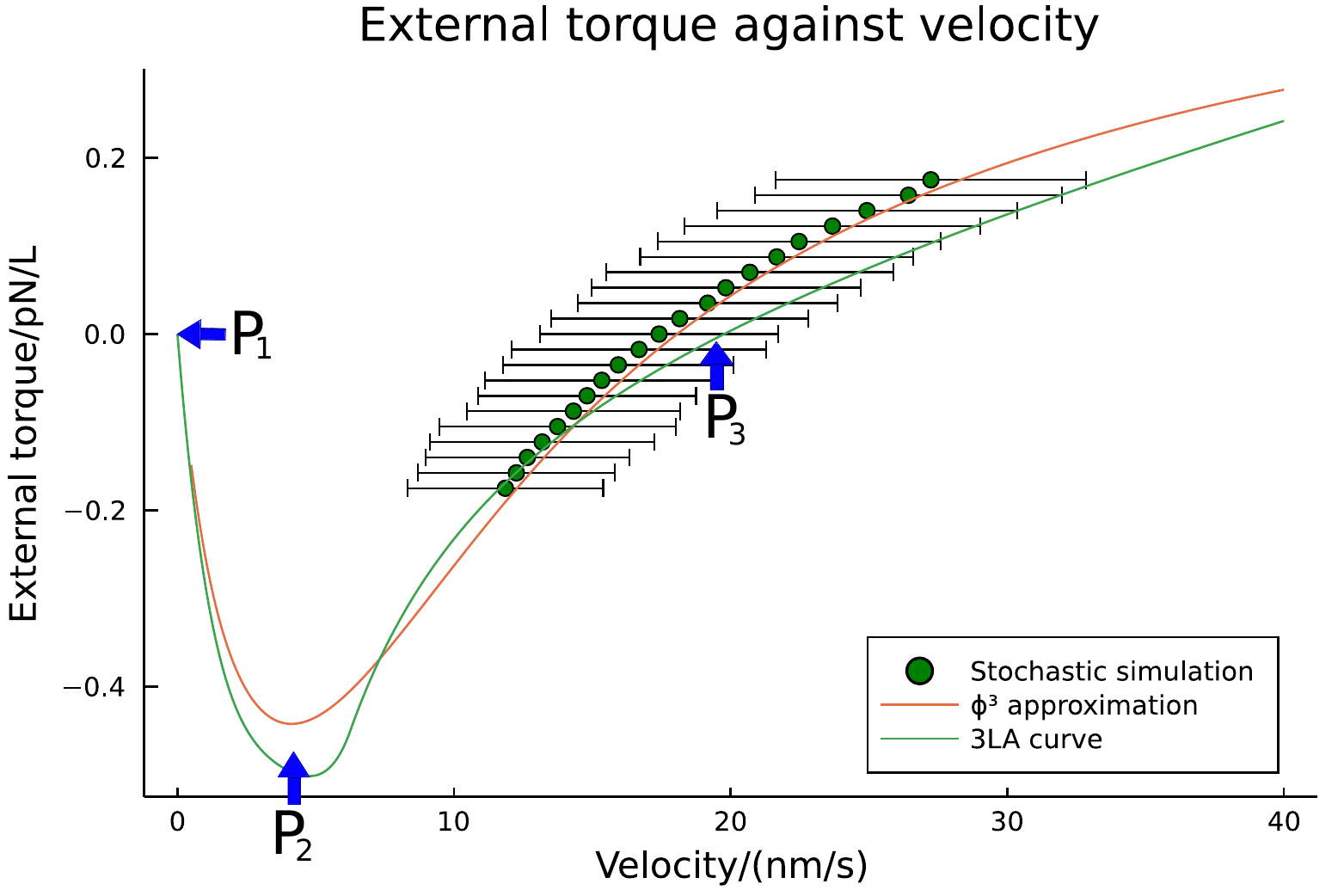} 
\caption{\label{fig:forcevel_stochast}
Torque-velocity relation obtained from the numerical solution of the continuum model (red curve),
from the approximate multi-line theory, Eq.~(\ref{eq:Motor-Curve}),
with $\beta$ from the perturbation theory (green curve),
and from stochastic simulations (symbols). 
Close to steady torque-free rolling, the agreement between numerics
and stochastic simulations is excellent; the approximate theory 
displays the correct overall shape.
The torque is given in pN$/L$ where $L$ is the virus' length
along the cylinder axis. 
For the stochastic simulations, the velocity has been averaged over the whole simulated 
trajectory with error bars displaying standard deviation. The points $P_1,P_2,P_3$ characterizing the "motor curve"
are discussed in the text.
}
\end{figure}

Fig.~\ref{fig:forcevel_stochast} shows the torque-velocity relation 
from a numerical solution of the continuum model (red curve)
as explained in section \ref{num_sec}, and the approximate multi-line theory
(green curve). The stochastic simulation results (symbols) shown additionally 
are explained in the next section.
In view of the approximations made, the green curve given by
Eq.~(\ref{eq:Motor-Curve}) captures the behavior
well on the semi-quantitative level and displays the same 
features as the numerically obtained red curve, namely three characteristic points:

($P_1$): for $\left(\omega,m\right)=\left(0,0\right)$
the virus is in the immobile state, which always should be a solution.
($P_2$): the point 
$\left(\omega,m\right)\approx\left(\frac{\sqrt{5}}{2}\omega_0,\left(5.6\,A-0.8\right)m_{c}\right)$ 
characterizes the maximum sustainable torque (in the deterministic, mean-field model). 
For counter torques of larger amplitude, the motion is no longer stable.
And finally ($P_3$): for 
$\left(\omega,m\right)=\left(\frac{\omega_0}{\sqrt{5A}},0\right)\propto\left(A^{1/2},0\right)$
the system is in the steady, self-propelled rolling state and torque free,
a behavior that is well captured already by Eq.~(\ref{om_virus}).

Concerning ($P_2$), interestingly for small activity $A\rightarrow0$ the maximum 
torque  the virus can sustain becomes independent of the activity
parameter: $m\to-0.8m_{c}$. The corresponding minimal speed under
subcritical forcing 
vanishes linearly with 
activity as $\omega=\frac{\sqrt{5}}{2}\omega_{0}\propto A\rightarrow0$.
In the same (singular) limit, $A\rightarrow0^{+}$, the torque-free
motion, i.e.~point ($P_3$), has no threshold and scales as $\omega_{free}\propto A^{1/2}\propto\phi_{c}\sqrt{\alpha V_{cut}}$,
as already derived previously. 
Thus, in the absence of external torque and for arbitrary low activity
$A$, one always has an (arbitrary slow) self-sustained rolling motion.

\section{Stochastic simulations}

\label{stochastic}

\subsection{Implementation and comparison stochastic vs.~continuum model}
\label{Gillespie_comp}

So far we investigated the problem of rolling on the mean-field level,
where we could show the existence of a steady rolling state for a virus having enzymatic NA activity. 
The questions whether this state is stable in view of the strongly fluctuating conditions 
at the nano scale and how the virus actually reaches this state, i.e.~self-polarizes,
demand a stochastic modeling framework. We here briefly explain the implementation
of the model using the Gillespie algorithm (with details given in appendix \ref{appstoch}) 
and then critically compare the stochastic simulations to the continuum theory.

Let us  consider a cylindrical virus, cf.~Fig.~\ref{figsketch},
of length $L$ and presenting a number of $N_{vir}\propto L$ discrete binding sites
and project the cylinder onto a single cross-section.
Then all binding sites (that one can assume randomly densely packed along the virus surface) get projected onto this circle.
As the contact zone size, $2 \phi_c R$ is fixed, 
the relation $N_{vir}\Delta x=2\phi_c R$ defines the 
effective size $\Delta x$ of a projected binding site.  
$N_{vir}\rightarrow\infty$ (i.e. $\Delta x\rightarrow 0$)  represents the continuum deterministic limit. 
In turn, small $N_{vir}$ can be interpreted as a circular virus, 
for which, using the size of an HA and a circular contact area,
one estimates $N_{vir}\simeq20$.

In the numerical algorithm,
space is discretized by $\Delta x$ on a large box 
(having typically $N=2000\gg N_{vir}$ sites) with periodic boundary conditions.
The angular variable used in the mean field model is discretized accordingly,
i.e.~by $\phi=2\phi_c n/N_{vir}$
defining the discrete bound linkers and free glycans,  $B=B[n]$ and $G=G[n]$, respectively. 
To obtain the effective molar concentrations 
of the spike proteins and the ligands one needs to convert from known surface densities on the virus and the substrate 
to volume densities.
Estimating the charateristic volume of a molecule
by $V_m\sim(10\,{\rm nm})^3$, 
the molar concentration is given by $C_M=\frac{1l}{N_A V_m}\simeq 1\,{\rm mM}$,
or in other words, 1 molecule$/V_m$ corresponds to $\simeq 1\,{\rm mM}$.

The Gillespie method \cite{Gillespie} is an event-driven  algorithm,
and time $t_i=\sum_{m=1}^i\Delta t_m$ is discretized in waiting times $\Delta t_m$. 
For every waiting time $\Delta t_m$ one considers all possible events,
i.e.~binding, unbinding and cutting with their discretized rates.
The waiting times are drawn according to
$ \Delta t_{m} = -\frac{\ln\xi}{a_T}$
where $\xi\in[0,1]$ is a uniform random variable
and $a_T$ is the sum of the rates of all possible events (see appendix \ref{appstoch} for details).
A second random number is used to choose which event takes place.  

Finally, one has to evaluate the torque balance using the newly obtained linker configuration
to determine the new center of mass position of the virus, $s(t_{i+1})$. 
Torque balance is assumed to be instantaneously established and,
expressing the contact interval $[-\phi_c,\phi_c]$ via the binding sites 
$[n_{L},n_{R}]$, the discretized torque balance  reads
\begin{equation}\label{torquediscretized}
\sum_{n=n_{L}}^{n_{R}}\left(n-s\right)^{3}\,B[n]  =0.
\end{equation}
This is a cubic equation for $s=s(t_{i+1})$ (for details see appendix \ref{appstoch})
and always has a real solution
that can be determined by root finding after every event that changes $B[n]$.
Having the new position $s$ (the nearest binding site is chosen), we get the  
new binding interval by shifting $B$ to its new center of mass position $s$.
This simulation yields trajectories $s(t_i)$, from which
angular velocities $\omega(t_i)$ and  velocities $v(t_i)=R\omega(t_i)$ can be readily determined, 
as well as the profiles $B[n], G[n]$ at every $t_i$.

To compare the stochastic implementation 
to the mean-field, continuum model
we first choose a high number of linkers, $N_{vir}=200$, such that
effects like stochastic reversals of the direction of motion or even complete detachment
of the virus do not occur (see next section for a discussion of these effects).
If not stated otherwise, we simulate the system for the  parameters,
already given and discussed in section \ref{basic}.
Going back to Fig.~\ref{fig:forcevel_stochast} that compares the torque-velocity relation 
obtained from stochastic simulations (symbols) to those from the numerical solution (red curve)
and the approximate multi-line theory (green curve) of the mean-field model, 
one can see that the numerics and the stochastic simulations agree very well
in a substantial region around the torque-free rolling.

Interestingly, in the stochastic simulations we were unable to get the full branch from 
the torque-free rolling down to the minimum of $m(\omega)$: 
the virus always reverted its direction for large counter-torque,
i.e.~the system jumped to the negative velocity branch -- note that for negative velocity,
the torque velocity is the same curve as displayed for positive velocities, but upside-down.
Without stochasticity, one would expect the branch to be stable down to the minimum
and only the branch from the minmum up to $(m,\omega)=(0,0)$ to be unstable.
The latter branch corresponds to bound linker distributions with a three-line profile,
i.e.~where the glycan is completely cut at the back.
This suggests that these states/profiles are important to understand the full
torque-angular velocity relation, but that they are dynamically unstable, 
especially in the presence of stochastic fluctuations.

\begin{figure}
\centering \includegraphics[width=1\linewidth]{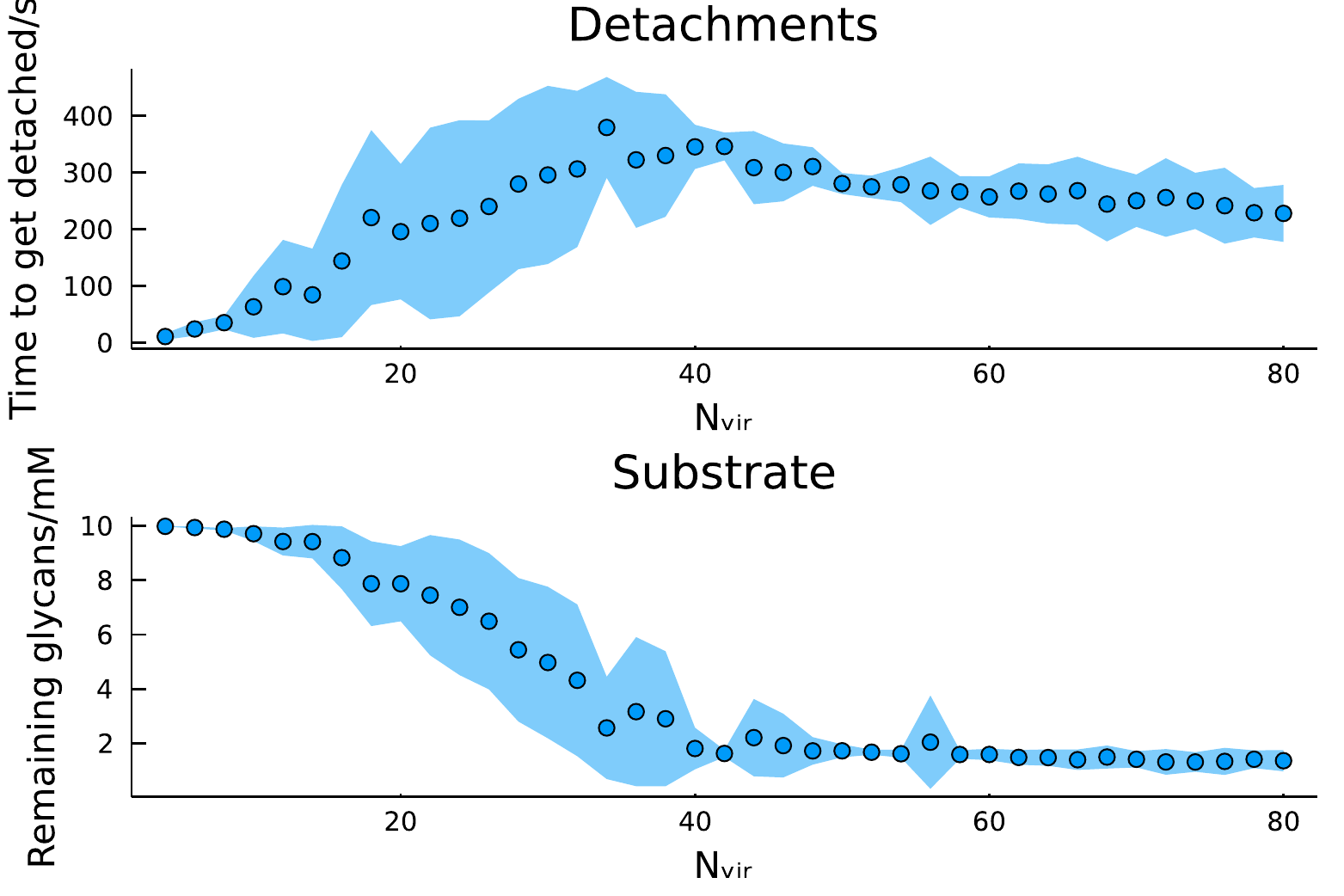} 
\caption{Study of a virus (radius 50 nm, $N_{vir}$ binding sites) rolling on a substrate 
with periodic boundary conditions (total number of binding sites 2000). 
Panel a) shows the average time until
the virus detaches completely from the substrate as a function of the number of linkers $N_{vir}$.
Panel b) shows the average glycan remaining on the substrate 
after  the virus has detached, again as a function of the number of linkers.
Averages were taken over 20 realizations for every value of $N_{vir}$,
shaded areas show the standard deviation.}
\label{fig:detachment} 
\end{figure}

\subsection{Stochastic effects: virus detachments and reversals}
\label{detach_reverse}

\begin{figure*}
\centering \includegraphics[width=.8\linewidth]{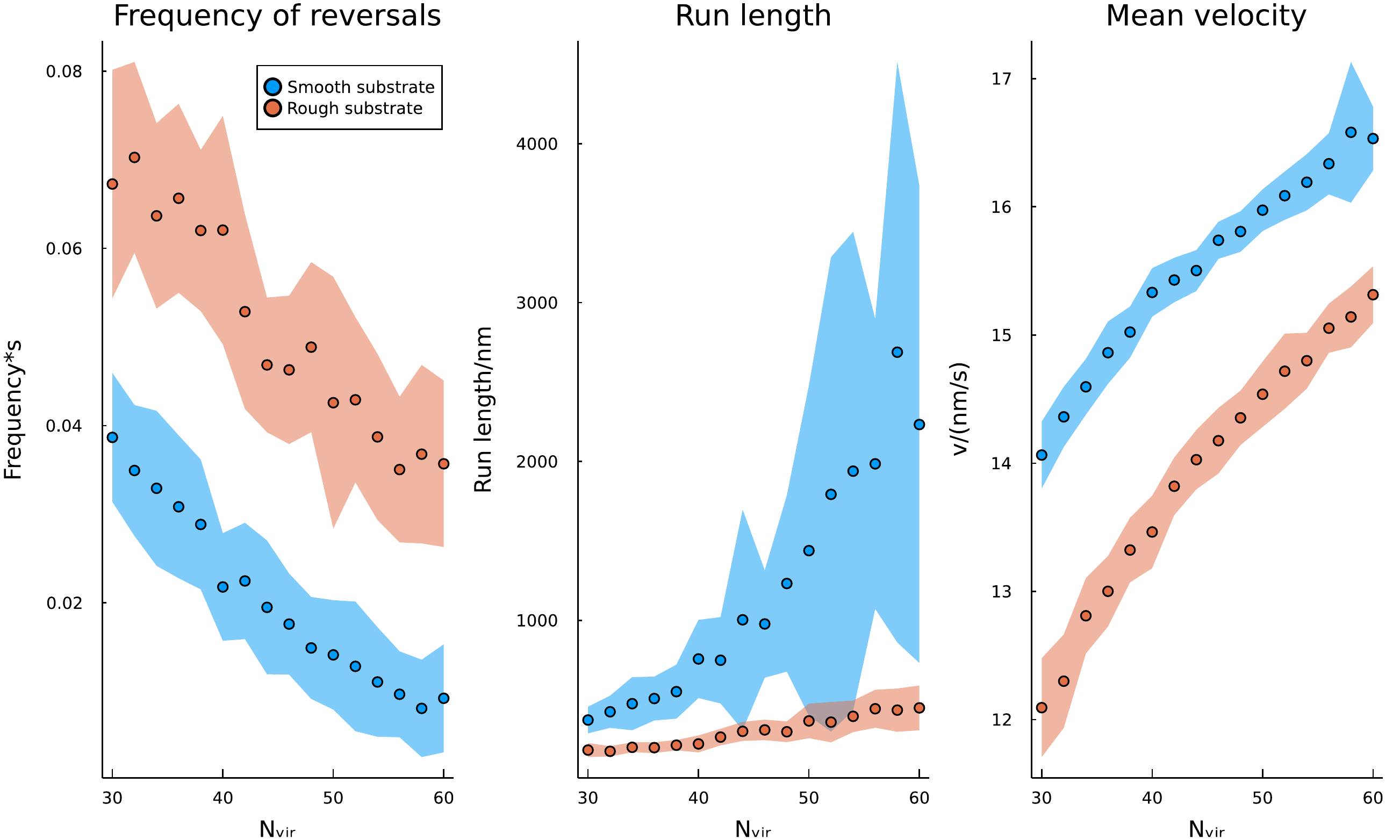}
\caption{Study of a virus (radius 50 nm, $N_{vir}$ binding sites) rolling on a substrate 
with periodic boundary conditions (total number of binding sites 2000) 
and infinitely rapid glycan recovery outside the virus' contact zone.
Panel a) shows the frequency of reversals (number per time) and
panel b) the average run length between two reversals.  
Panel c) gives the average velocity obtained from the two former quantities. 
For the blue data the glycan distribution on the substrate was perfectly homogeneous
at the standard value used, $G_{0}=10$, 
while for the red data a noisy distribution was implemented (with strong noise amplitude of $G_{0}/2$). 
All averages were taken over 20 runs of 500 s running time each,
shaded areas show the standard deviation.
}
\label{fig:VFR} 
\end{figure*}

The stochastic implementation in addition allows to investigate the effects of finite linker numbers
on the overall behavior of virus rolling, inducing for instance reversal of rolling direction
or even complete detachment.
Decreasing $N_{vir}$ to values below 100 (for the given, realistic parameters),
Fig.~\ref{fig:detachment}a) shows the time it takes for a virus, initially
placed on a homogeneous glycan-covered substrate, to get completely detached
-- i.e.~reaching the state with $B[n]=0$ for all $n$ in the contact zone. 
In turn, Fig.~\ref{fig:detachment}b) shows the averaged amount of glycan 
that is still left on the substrate, after the virus has detached.

Fig.~\ref{fig:detachment} suggests that the origin of detachment for small values of $N_{vir}$
is due to intrinsic stochasticity of binding, as the glycan level after detachment is still high.
In contrast, for large values of $N_{vir}$ the virus rather detaches
since the glycan level becomes low. The latter is due to the fact that the virus 
consumes part of the substrate while rolling. It may also change direction of movement 
or go through the periodic box such that it crosses again the same, already partially 
glycan-depleted region more than once.

If the substrate were not consumed one would expect the time to get detached 
to grow exponentially with the number of linkers, 
since so does the number of possible configurations of $B[n]$. 
In order to study reversals and run times in a way that is not influenced by the history of the virus' path,
in the following we will consider the case were the glycans are recovered by the cell.
For simplicity we assume this recovery to be infinitely fast.
Specifically, whenever a virus has rolled over a part of the substrate, 
the respective value $G(n)$ is restored to the initial level $G_0$.
This corresponds to a virus rolling on a cell that has a high membrane diffusivity and rapidly 
reshuffles its surface glycans. We also come back to this point in section \ref{sec_discussion};
note that glycan reshuffling via lateral diffusion within the cell membrane has been readily 
observed in experiments \cite{GlycanDiffusion1,GlycanDiffusion2}. 

Implementing this infinitely fast glycan recovery outside of the contact zone of the virus
one can study the statistitics of the reversals of direction of the rolling virus
and quantify the typical runlengths as a function of parameters and independent of
the simulation box size. The results are shown in Fig.~\ref{fig:VFR}.
The blue data in Panel a) show the frequency of reversals as a function of the number of linkers
for homogeneous glycan distribution $G_0$ outside the contact zone.
For the red data, we added a stochastic glycan distribution (noise level $50\%$ of mean value).
In both cases, the reversals decrease if the linker number increases. 
For the blue data, these reversals are solely due to the intrinsic stochasticity of the dynamics.
As expected, for the noisy glycan distribution, reversals occur more frequently,
but the rolling is still robust.  

Fig.~\ref{fig:VFR}b) shows the average runlengths, i.e.~the distance traveled 
between two reversals, obtained from the same raw data. 
The runlength increases with the number of linkers and can easily reach several microns
(meaning many tens the virus diameters) for a homogeneous glycan distribution.
Noisy glycan distributions impede this substantially, but the virus still travels few times 
its size for still moderate linker numbers. 
Finally, from panels a) and b) one can estimate the mean velocity of the virus 
as shown in Fig.~\ref{fig:VFR}c). 
This is in good qualitative agreement with the theory/continuum model,
which for the given parameters (and $N_{vir}\rightarrow\infty$) is 20 nm/s,
cf.~section \ref{ana_discussion}.

\section{Further topics} \label{further topics}

\subsection{Approximations made, especially detailed balance for on-off kinetics}
\label{detailed_balance}

Within the simple model approach it is very satisfying
to see the good agreement between the numerical and analytical
approaches to the mean field model and the stochastic implementation.
However, to be amenable to an analytical treatment, 
we made several -- in part strong -- approximations that should be critically discussed.

First, in section \ref{sectorquebal}
concerning the torque balance, we applied a small angle approximation
for the contact angle. We checked numerically that for the given, realistic parameters,
this has only a minor quantitative effect. 
Second, we neglected the effect of linker compression. 
In small angle approximation, linker compression can be included even
analytically, with details given in appendix \ref{compression}.
Again, this  only leads to a quantitative correction.

The most critical approximation was made in section \ref{dynamics}: 
namely, the simple model  only approximately fulfills 
detailed balance for the HA-glycan on-off kinetics.
In fact, while in section \ref{contact_interval} we used the force-dependence 
(and hence angle-dependence) 
of the  attachment/detachment kinetics of the linkers to determine the size of the contact interval,
this dependence was neglected in the dynamic equations, 
Eqs.~(\ref{eq:ViroBoidfinB}), (\ref{eq:ViroBoidfinG}). 

To improve on this point we now assume that the dissociation constant increases with 
the ``Boltzmann factor'' of the elastic stretch energy
\beq\label{Kdphi}
\frac{K_d(\phi)}{K_d(0)}=\exp\left(\frac{E_{el}}{k_BT}\right)
=\exp\left(\frac{SR^2}{8k_BT}\phi^4\right).
\eeq
The equilibrium probablity distribution of the bound linkers is then given by 
$B_{pl}(\phi)=\frac{C_0-C_1}{2}$, as given in section \ref{analytic_forced_roll},
but now with the angle-dependent $K_d$ 
entering $C_0$, $C_1$, explicitly
\beq\label{pbound}
\frac{B_{eq}(\phi)}{H_0}=
\frac{1}{2}\left(\bar{B}-\sqrt{\bar{B}^2-4\frac{G_0}{H_0}}\right)
\eeq
with $\bar{B}=1+\frac{G_0+K_d(\phi)}{H_0}$.
The resulting $B_{eq}(\phi)$ implies a bell-shaped linker profile, with a maximum at $\phi=0$
and decaying rapidly (like $\exp(-\phi^4)$)  for finite angles. 

We can now determine the size of the contact interval more properly:
defining $\phi_c$ as the point where $p_{\rm bound}=
\frac{B_{eq}(\phi)}{H_0}$ drops below $1/2$,
from Eq.~(\ref{pbound}) one gets 
$G_0-\frac{H_0}{2}=K_d(\phi_c)$.
For typical parameters, $H_0/2\ll G_0$ holds. Inserting Eq.~(\ref{Kdphi}) and solving for $\phi_c$ 
then yields exactly the scaling result, Eq.~(\ref{contact_angle}).

\begin{figure}[t]
\centering 
\includegraphics[width=.9\linewidth]{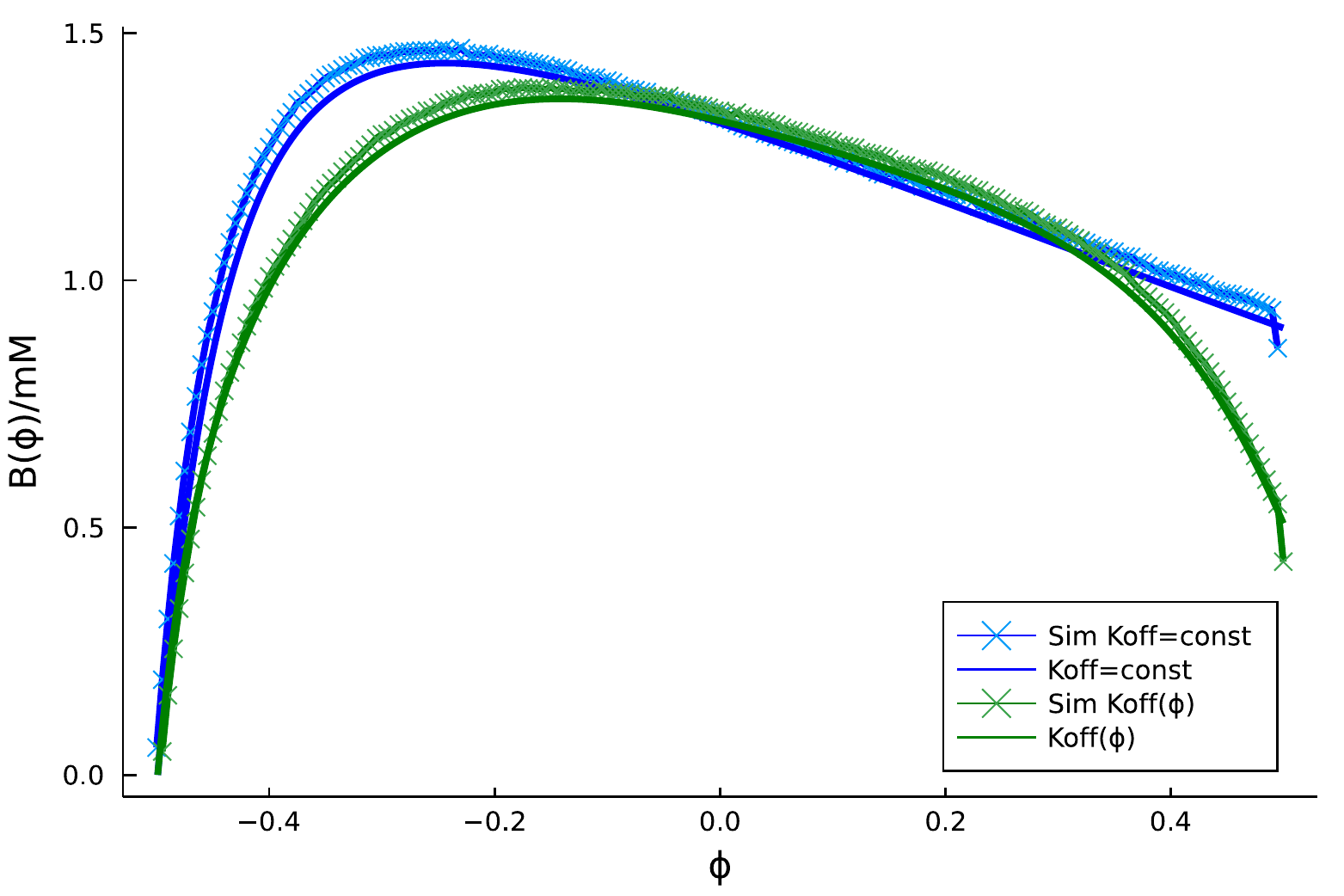} 
\caption{\label{fig:Bwandwo}
Test of the effect of detailed balance of the on-off kinetics. 
Shown are bound linker distributions
obtained numerically (solid) and by stochastic simulations (symbols). 
The blue curves are for the simple model (discussed throughout so far, 
without detailed balance),
while the green curves obey detailed balance, i.e.~$K_d(\phi)$ is angle-dependent
as given by Eq.~(\ref{Kdphi}).
In the latter case, the sloped linker profile in the center prevails 
and the virus is still able to roll steadily.
Stochastic simulations are averaged over $10^6$ realizations.}
\end{figure}

To scrutinize the effect of the detailed balance of the on-off kinetics on
the bound linker distribution and ultimately on the rolling motion,
we compared numerical solutions of the mean field model
to stochastic simualtions, both without detailed balance (constant $K_d$)
and with detailed balance, i.e.~$K_d(\phi)$.
The result is shown in Fig.~\ref{fig:Bwandwo}.
The blue curve shows the case studied before, the two slopes 
quantified by $\alpha$ and $\beta$ being clearly visible.
The green curve shows the case with detailed balance. One can
see that the bound linker profile decays much stronger both in rolling direction
(to the left) and at the trailing edge (to the right), but overall
the sloped distribution in the center region prevails and is sufficient
to allow persistent rolling motion.   
We also checked that the speed is only quantitatively affected
(for the given, realistic parameters by 10-20\%).

\subsection{Linker distribution as an "internal flywheel"}
\label{flywheel}

As observed already in Ref.~\cite{virusPRL}, the Gillespie simulation-based 
computer experiments reveal a surprising robustness of the rolling virus: 
even when facing obstacles in form 
of glycan-depleted spots on the surface, the virus often just rolls over them as if it 
possesses an internal "inertia". 
A once directionally polarized virus can even persistently roll against a glycan gradient as long as there is a sufficient amount of glycan left to bind. Thus the virus is not performing a (chemo)taxis on the glycan concentration (as one would expect for 
a burned bridge Brownian ratchet). Instead the glycan acts here as a mechano-chemical free-energy source in analogy to the role of ATP for classical molecular motors. The fact that the glycan is confined (or sometimes even immobilized) on a 2D surface while ATP typically freely diffuses in 3D is only a superficial difference.

To understand the origin of the observed processivity behavior , we pose the 
following question: What happens when a steady rolling virus at initial  
angular velocity $\omega_{free}$ 
is suddenly stopped -- by an external force, such as an immovable obstacle -- 
at time  $t=0$? How does its torque $m(t)$ dynamically respond to 
this sudden constraint?

\begin{figure}[t]
\centering 
\includegraphics[width=\linewidth]{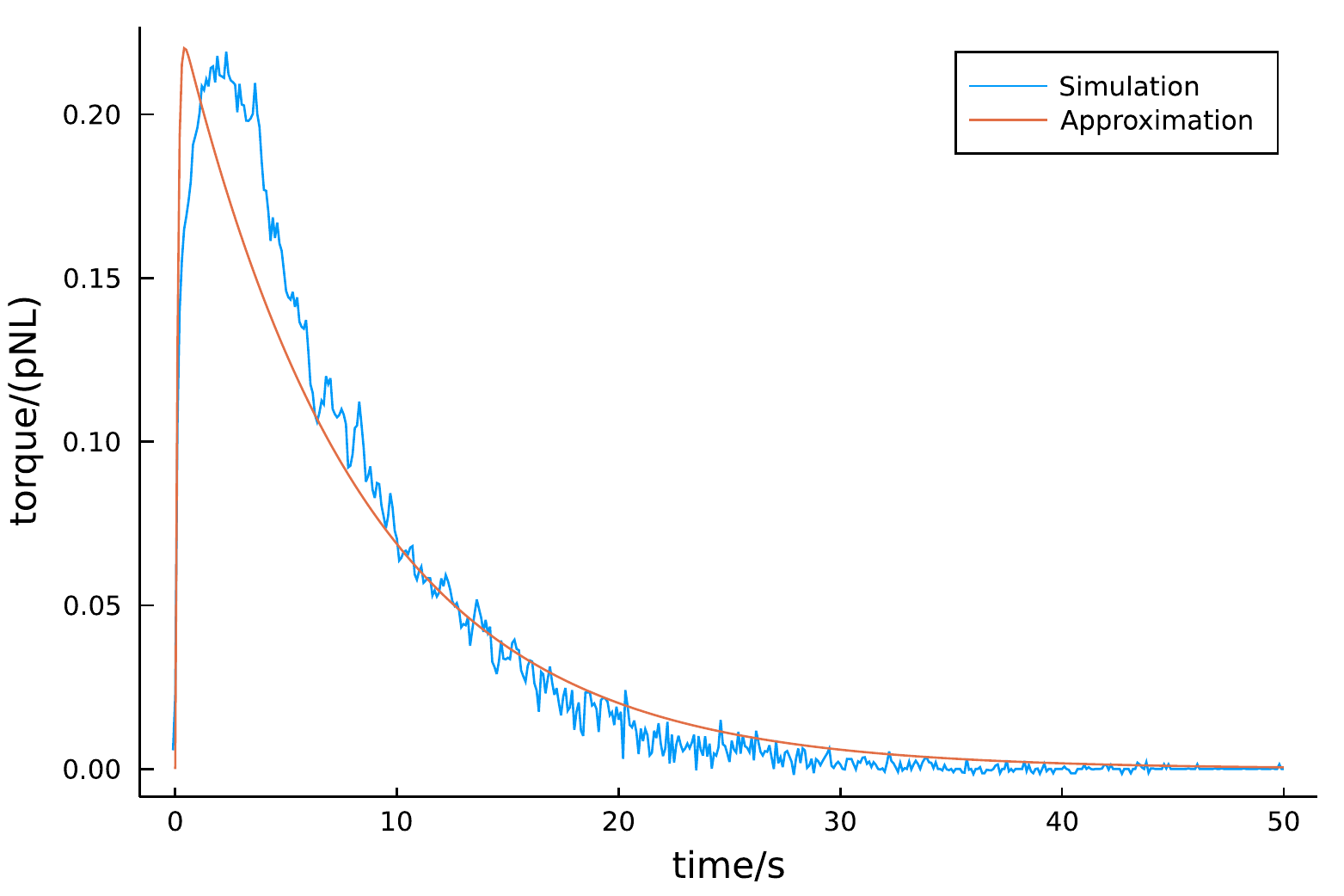} 
\caption{\label{fig:flywheel}
Demonstration of the ``mechano-chemical flywheel'' effect.
A virus that was rolling steadily was suddenly stopped at $t=0$. 
Since the internal linker profile needs time to relax, a characteristic buildup
of torque occurs with a characteristic time of the order of $10\,$s.
Shown are results from stochastic simulations (blue; averaged over 10 events)
and the scaling estimate, Eq.~(\ref{flywheeleq}), in red (with no fitting parameter).
Linker stiffness $S=0.1k_BT/{\rm nm}^2$, other parameters as given in section \ref{basic}.
}
\end{figure}

Initially, for $t<0$, the virus was rolling in a torque-free state, implying that any torque buildup 
starts from $m(0)=0$. The virus being blocked for $t>0$, the advective (rolling) term is now missing 
in Eqs.~(\ref{eq:ViroBoidfinB}), (\ref{eq:ViroBoidfinG})
and the linker distribution in the front region will  respond by rapidly equilibrating 
to the plateau-value there. This front equilibration 
has the timscale $t\sim B_{pl}/\alpha$ and due to torque imbalance 
it is accompanied by an buildup of torque $m(t)$. 
While the front region's slope (cf.~Fig.~\ref{fig:VelRegimes}) 
becomes flatter during this fast and transient process, 
the more extended, cutting-dominated rear region stays roughly unperturbed
as it responds much slower. 
This implies a maximum torque $m_{max}\propto - \beta / \omega_{free}$ 
given only by the cutting-induced gradient at the rear. 
Finally, on the long timescale $t\sim B_{pl}/\beta$, 
the progressive NA linker-cutting makes
the linker profile decay to zero in the whole contact zone, 
leading to a gradually vanishing torque $m(t) \rightarrow 0$.  
In summary, a suddenly stopped virus responds with a transient torque buildup 
for $t>0$ of the approximate form 
\begin{equation}\label{flywheeleq}
m(t)\simeq m_{max} \tanh\left(\frac{\alpha}{B_{pl}} t\right) e^{-\frac{\beta}{B_{pl}} t}
\end{equation}
where $m_{max}\simeq p_{act}/\omega_{free}$.

Figure \ref{fig:flywheel} shows a comparison of the torque buildup measured 
in stochastic simulations (averaged over 10 stopping events) and the scaling
formula, Eq.~(\ref{flywheeleq}). It confirms the interpretation 
that a stopped virus mobilizes its bound linkers and puts up dynamic resistance 
against the obstacle. One could say that the system formally behaves 
as if it possessed a built-in "flywheel" -- here of mechano-chemical origin --
that is dissipatively coupled to the rolling angle variable and 
tends to maintain its angular momentum.  
This "flywheel" dynamic response is associated to the re-equilibration of
an internal nonequilibrium steady state. It seems 
to be a defining feature of other dissipative rolling objects as well \cite{DNAmw1,Baumann,Bazir} 
and deserves closer inspection in the future.

\section{Discussion}
\label{sec_discussion}
We have proposed and analyzed a new mechanism allowing a virus to actively roll along 
a substrate and shown that  this rolling is essentially inevitable, 
once few basic conditions are met: namely, 
(1) existence of a rolling axis for the virus,
(2) presence of ligands on the substrate, 
and (3) viral spikes (one or several types) that bind 
{\it and} enzymatically cut these ligands.
The obvious question to ask next is: how does the virus benefit from it? 
\subsection*{Why the rolling?}

Obviously, being a nanoscale object the virus could just move through the bulk of a 
low viscosity fluid via thermal diffusion. However, 
in the case of mucus-binding viruses like influenza, the large viscosity and 
the gel-like nature of 
the environment severely limits this. 
An additional danger influenza faces is getting stuck in/to the mucus and 
eventually being swept away by the beating cilia of our respiratory tract.
To make the situation even worse for the virus, 
during its entry phase \cite{Boulant,Alsteens_virusentry}
the virus can become localized at wrong spots on the cell membrane
(or on a wrong cellular structure like a cilium), being unable to enter the cell at all. 
If the cell membrane is sufficiently fluid, two-dimensional (2D) diffusion 
can help out, but not if the glycan receptors (glyco-lipids and proteins) are 
trapped within local membrane domains controlled by the cytoskeleton underneath.  
Influenza could respond by weakening its bonds with the immobile glycans 
in order to still diffuse on the quasi-rigid substrate, the drawback being 
that increasing diffusivity also increases the virus-membrane detachment rate. 
All these are likely reasons for influenza having evolutionary come up 
with the active enzyme solution.

In spite of sufficiently strong collective, multi-linker 
binding by the HA-spikes, the virus can still actively weaken these links 
via the NA-spikes' cutting activity. However, if this form of "stick-and-cut" behavior 
was  spatially uncoordinated, it would be difficult to comprehend how 
it could give rise to any sufficiently fast or even  directed motion. 
How should the virus decide which way to go and keep a certain persistence of direction?
Obviously, the (immobile) glycan-covered substrate could help the virus to some extent
by keeping records of where the virus has already been: the virus would then simply 
statistically avoid the NA-generated glycan-depleted regions and perform a form of 2D 
self-avoiding walk, which is the idea underlying the so-called 
burnt-bridge Brownian ratchet \cite{Blumen,Krapivsky}. 
This strategy could be  enhanced by a polarized distribution of the spikes \cite{Fletcher},
with  NA enriched at the rear, depleting more glycans there, and HA enriched in front,
both effects enhancing forward binding and stabilizing directional movement. 
Indeed this is  observed for IVA where HA and NA are  physically separated molecules
``floating'' rather freely in the membrane, where they can polarize 
the virus via a partial phase separation. The motion in this case happens 
{\it along the long axis} of the ellipsoidal or filamentous virus. 
However, for IVC the two spikes are ``glued together'' into a single, 
inseparable unit, the HEF protein \cite{HEFref}. Naively, this 
additional constraint would make the virus less motile, 
yet the opposite is true: IVC moves about 5-10 times faster than IVA 
\cite{Sakai_Saito_IVA,Sakai_Saito_IVC}. 
In addition to its larger speed, most notably IVC  moves {\it orthogonally to its long axis}. 

It seems that these two propulsion modes -- parallel and orthogonal to the axis -- 
have radically different physical mechanisms. Here we suggested that the motion orthogonal 
to the cylindrical axis is tightly 
coupled to axial rotation. This coupling
is more than just an easy, low dissipation mode of motion, but rather intrinsically linked 
to the very mechanism of dynamic linker polarization: on the one hand, the angular rotation 
itself leads  to a linker polarization within the virus-substrate contact zone. 
On the other hand, the linker polarization front vs.~rear 
gives rise to a torque and the angular velocity. 
Both 
interdependent effects are inseparable and make rolling propulsion
fast, robust and efficient compared to longitudinal gliding.

The emergence of rolling can be seen as a step in the evolutionary race between the virus 
and the host. Naturally, this race is still ongoing and the host could also take counter measures. 
One possibility is the mixing-up and quick replacement of the cut glycans by fresh ones
(cf.~the glycan recovery  in section \ref{detach_reverse}). 
However, for this to be efficient it has to affect the contact zone, 
as the rolling virus is not very sensitive to gradients outside. 
Hence glycan replacement must be extremely fast (the virus passes the contact zone within few seconds). 
Another possibilty is to "clog" the rotation by firmly binding ligands (antibodies) 
to some of the free HA/HEF. In this case, the virus would have to resort 
to other, less efficient propulsion mechanisms like gliding. Alternatively, 
it could counteract the clogging by allowing sterically blocked HA/HEF 
to float by keeping the transmembrane proteins (including the spikes) in a fluid state. 
As just sketched here, the game-theory of this intricate evolutionary 
race promises new surprises for future studies.  

\subsection*{ Experimental Questions}

From the experimental point of view, advancing elegant in vitro setups like those 
by Sakai et al. \cite{Sakai_Saito_IVC,Sakai_Saito_IVA} should allow to quantitatively probe 
the detailed mechanism proposed here, the most robust and easy-to-test predictions being:\\
(1) The rolling speed as a function  of  NA,  HA, and glycan surface-concentrations.  
Especially  the characteristic square root relation $\omega \propto \sqrt{V_{cut}}$,
cf.~Eq.~(\ref{om_virus}), between the angular velocity and the cutting rate $V_{cut}\propto [NA]$.\\
(2) The force-velocity or ``motor relation" shown in Fig.~\ref{fig:forcevel_stochast} 
and the characteristic points $P_1$-$P_3$ therein, as well as the scaling behavior of the curve.\\
(3) The instantaneous force response of a stalled, immobilized virus, 
as shown in Fig.~\ref{fig:flywheel} and quantified in Eq.~(\ref{flywheeleq}).\\

Experimentally these should be accessible 
as the filamentous viruses are large (long axis one micron) and sufficiently slow (tens of nm/s) 
to be readily captured by various microscopy methods. Even  smaller viruses
have been studied already using high resolution techniques \cite{Sandoghdar}. 
The force magnitudes (tens of pN) and their moderately slow time evolution (few seconds) 
are also well within the range of common 
force spectroscopy methods 
\cite{Hermann1,Hermann2}. 
 To test the theory it is most practical to utilize filamentous viruses. 
Although the model developed here also applies to spherical viruses on small time scales, 
the  high rotational diffusion constant of a tiny nanosphere and the small number of 
 linkers making contact ($N_{vir}\simeq20$) 
will in this case give rise to large  orientational and velocity
fluctuations. As for any directionally self-propelled object of characteristic size $L$, 
there will be a crossover from ballistic to diffusive motion at a 
timescale $t_{rot}\propto 1/D_{rot}$ with $D_{rot}\propto L^{-3}$ 
the rotational diffusion constant. While for a sphere $L$ will be the radius,
for a cylinder it will be its length, making the directed propulsion of the  latter experimentally much easier to probe.

Many interesting questions still remain, including the role of the different motility modes -- gliding by spike polarization \cite{Fletcher}
	and rolling \cite{Sakai_Saito_IVC,Sakai_Saito_IVA} -- under similar conditions. Can IVA actually
	switch between the two motility modes? And how do IVs finally switch from rolling to entering the cell
	\cite{Boulant,Alsteens_virusentry,Du_HA-NA}?
	
	On the physics side there is an interesting conceptual question that still needs to be addressed: 
	Other non-equilibrium filamentous rollers are known to exist, 
	but these are bulk-driven, meaning there is a matter-energy flow (e.g. of heat \cite{Baumann} or solvent \cite{Bazir})
	through the cross-section of the cylinder.
	So what are similarities and differences of active rollers that are bulk-driven vs.~surface-driven
	in view of a general description, their efficiency, or the existence and nature of their internal flywheel modes?

\section{Conclusions}

In conclusion, the rolling mechanism renders influenza an even 
smarter adversary than previously thought.  
In contrast to classical virology dogmas, this virus type displays a 
``surface metabolism'' harvesting the chemical energy of sugars on the host's membrane for directional force generation.
In turn, this transforms the whole virus capsid into a complex rotary motor.
The underlying linker dynamics shares similarities 
with passive  (e.g.~shear induced \cite{Hammer04, KornSChwarzPRE, Anil})
or other \cite{Korosec_Emberly} adhesive rolling mechanisms, 
as well as with collective motor ensembles  \cite{JulicherProstPRL,JulicherRMP,Reimann},
but with the crucial addition of the enzymatic substrate cutting.
The mechanism discussed here for influenza 
should also apply to other viruses having enzymatic spike proteins: 
candidates are the toro-virus and some of the beta-corona-viruses \cite{HEFinCoronaToroV}.
We can also flip the coin and learn from the virus' workings: 
in fact, the combination of a binding molecule, a cutting enzyme and 
the spherical/cylindrical geometry has been already used to 
propel DNA-coated beads \cite{DNAmw1,DNAmw2,DNArods} along RNA-covered surfaces
and the insights developed here could now be used to optimize such and related artificial rollers.

\section{Appendix}
\appendix

\section{Calculation in angle space and scaling}
\label{angle_space_scaling}

The calculation of the torque becomes especially transparent in angle space, where
the two-line linker distribution reads
\begin{equation}\label{twolineangle}
B\left(\phi\right)=\begin{cases}
\frac{\alpha}{\omega}(\phi+\phi_{c}) & \text{ for \ensuremath{-\phi_{c}\le\phi\le\phi_{pl}}}\\
\frac{\alpha}{\omega}(\phi_{pl}+\phi_{c})-\frac{\beta}{\omega}\left(\phi-\phi_{pl}\right) 
& \text{ for \ensuremath{\phi_{pl}\le\phi\le\phi_{c}}}
\end{cases}
\end{equation}

Rewriting the torque, Eq.~(\ref{eq:Torque m}), by again using
that a constant linker distribution is torque-free, yields
\begin{equation}
m=-\frac{m_{0}}{H_{0}}\frac{1}{\omega}\left[\alpha\int_{-\phi_{c}}^{\phi_{pl}}(\phi-\phi_{pl})\phi^{3}d\phi-\beta\int_{\phi_{pl}}^{\phi_{c}}(\phi-\phi_{pl})\phi^{3}d\phi\right]
\end{equation}
In the first integral, one substitutes $\tilde{\phi}=\phi+\phi_{c}$
and  uses that $\phi_{pl}+\phi_{c}\ll1$. In the second 
one uses $\phi_{pl}\simeq-\phi_{c}$. Both amount to the same statement,
that the plateau is rapidly reached (i.e.~closeby the boundary
in rolling direction). One obtains
\begin{equation}\label{phi_step}
m=-\frac{m_{0}}{H_{0}}\left[\frac{\alpha}{\omega}\frac{1}{2}\left(\phi_{pl}+\phi_{c}\right)^{2}\phi_{c}^{3}-\frac{\beta}{\omega}\frac{2}{5}\phi_{c}^{5}\right],
\end{equation}
and using $\phi_{pl}+\phi_{c}=\omega t_{m}$ one regains 
\begin{equation}
m=-\frac{m_{0}}{H_{0}}\left[\alpha\omega\frac{1}{2}\left(t_{m}\right)^{2}\phi_{c}^{3}-\frac{\beta}{\omega}\frac{2}{5}\phi_{c}^{5}\right].\label{forcevelscale}
\end{equation}

As stated already in section \ref{sec_scale} when discussing the scaling,
if the plateau is reached sufficiently rapidly, the torque integral
over the negative slope region is independent of the dynamics.  
Basically, there is degradation everywhere, resulting just in a power of $\phi_{c}$, 
the contact area size. It only remains the $1/\omega$-dependence 
from the slope in angle space. 
In contrast, for the first, passive contribution, cf.~Eq.~(\ref{phi_step}), the dynamics is important: 
the angle-integrated slope yields a square-dependence in the plateau angle and hence in $\omega$, 
which together with the $1/\omega$ from the slope yields a linear friction $\propto\omega$.

\section{Effect of linker compression}
\label{compression}

In section \ref{sectorquebal} we introduced a distribution of linker extensions,
that was assumed to be zero at the symmetry axis and growing quadratically
with angle $\phi$. This implied only tensile linker forces for
$\phi\neq0$ and a delta-peak counter-force with opposite sign at $\phi=0$. 
Physically, this assumption means that the stiff viral spike proteins 
can sustain a compressive load with little compliance but they easily stretch 
when tension is applied.

One can take the concept of linear linker elasticity 
more seriously and allow the virus to push down on and compress the linkers. 
Assuming a similar chain stretching as before, 
including  the possibility that the virus cross-section is shifted downwards by a 
(to be determined) length $l_{0}$, one has $l(\phi)=R(1-\cos(\phi))-l_{0}$.
The energy of the elastic spring foundation reads (for small $\phi$) 
\begin{eqnarray}
e_{el}(\phi)=\frac{S}{2}l^{2}(\phi)\simeq\frac{SR^{2}}{2}\left(\phi^{2}/2-\varepsilon_{0}\right)^{2},
\end{eqnarray}
with $\varepsilon_{0}=l_{0}/R$ the vertical deformation.
It is determined to be $\varepsilon_{0}=\frac{1}{6}\phi_{c}^{2}$
by the vanishing force condition, 
$\int_{-\phi_{c}}^{\phi_{c}}f_{el}(\phi)d\phi=0$,
with $f_{el}=-\frac{\partial e_{el}(\phi)}{\partial l}$.
The torque balance, Eq.~(\ref{eq:Torque m}), then generalizes to 
\begin{eqnarray}
m_{{\rm incl\,compr}}%
=-\frac{m_{0}}{H_{0}}\int_{-\phi_{c}}^{\phi_{c}}B(\phi)\left(\phi^{3}-2\varepsilon_{0}\phi\right)\,d\phi\,.
\label{newtorque}
\end{eqnarray}
In addition to the third moment, 
there now also is a contribution from the first moment, having 
opposite sign.
Evaluating this torque using Eq.~(\ref{twolineangle}) leads to 
$\omega_{{\rm incl\,compr}}=\sqrt{2/3}\,\omega$
implying a reduction of $\simeq20\%$.

\section{Details for the stochastic implementation}
\label{appstoch}

The Gillespie algorithm \cite{Gillespie} is event driven:
at any time step,
one decides whether at the picked position $n$ 
a binding event ($B[n]\rightarrow B[n]+1$ and $G[n]\rightarrow G[n]-1$),
an unbinding event  ($B[n]\rightarrow B[n]-1$ and $G[n]\rightarrow G[n]+1$) 
or a cutting event ($G[n]\rightarrow G[n]-1$) occurs. 
The weights of these events can be read off directly 
from Eqs.~(\ref{eq:ViroBoidfinB}), (\ref{eq:ViroBoidfinG}):
\begin{eqnarray}
&&a_b[n]=k_{on}G[n](H_0-B[n])\,,\,\,\,a_u[n]=k_{off}B[n], \nonumber \\
&&a_c[n]=V_{cut}\frac{G[n]}{G[n]+K_M},
\end{eqnarray}
which must be normalized by the sum $a_T=\sum_n (a_b[n]+a_u[n]+a_c[n])$ 
to assign probabilities to each process.  
The waiting times $\Delta t_{m}$ between two events
can then be picked 
as
$ \Delta t_{m} = -\frac{\ln\xi}{a_T}$,
with $\xi$ a uniform random variable on $[0,1]$.

Finally, to find the new center of mass position of the virus, 
at each step one solves Eq.~(\ref{torquediscretized}),
i.e.~the cubic equation,
\begin{equation}
s^{3}+c_{2}s^{2}+c_{1}s+c_{0}=0,
\end{equation}
with coefficients determined by the bound linkers as
\begin{eqnarray}
&&c_{2}=-\frac{3}{\Sigma}\sum_{i=i_{L}}^{i_{R}}n\,B[n]\,,\,\,\,
c_{1}=\frac{3}{\Sigma}\sum_{i=i_{L}}^{i_{R}}n^{2}\,B[n]\,,\nonumber\\
&&c_{0}=-\frac{1}{\Sigma}\sum_{i=i_{L}}^{i_{R}}n^{3}\,B[n]\,,
\end{eqnarray}
where $\Sigma=\sum_{i=i_{L}}^{i_{R}}B[n]$.

\end{document}